\documentclass{aastex631}
\usepackage{amssymb}
\usepackage[caption=false]{subfig}
\usepackage{graphics}
\usepackage{amsmath}
\usepackage{amsfonts}
\usepackage{bm}
\usepackage[]{latexsym}
\usepackage{cellspace}
\usepackage{mathtools}
\usepackage{amstext}
\usepackage{stmaryrd}
\usepackage{stackrel}
\usepackage{graphicx}
\usepackage{esint}
\usepackage[utf8]{inputenc}
\usepackage{blindtext}
\usepackage{float}
\restylefloat{table}
\usepackage{booktabs}
\usepackage{enumitem}
\usepackage{amssymb,amsbsy,epsfig,graphicx}
\usepackage{color}
\usepackage{etoolbox} 
\usepackage{lipsum} 
\usepackage{eufrak}
\usepackage{verbatim}
\usepackage{bm}
\usepackage{hyperref}
\usepackage[capitalize]{cleveref}
\hypersetup{pdfstartview=FitV,colorlinks=true,linkcolor=midblue,citecolor=midblue,filecolor=midblue,urlcolor=midblue}

\definecolor{midblue}{rgb}{0,0,0.5}

\def\actaa{\ref@jnl{Acta Astron.}}      

\renewcommand{\d}{\mbox{${\rm d}$}}

\newcommand{\be}{\begin{equation}}\newcommand{\ee}{\end{equation}}
\newcommand{\bea}{\begin{eqnarray}}\newcommand{\eea}{\end{eqnarray}}
\newcommand{\brr}{\begin{array}}\newcommand{\err}{\end{array}}
\newcommand{\bit}{\begin{itemize}}\newcommand{\eit}{\end{itemize}}
\newcommand{\ben}{\begin{enumerate}}\newcommand{\een}{\end{enumerate}}

\newcommand{\ba}{\begin{array}}
\newcommand{\ea}{\end{array}}

\begin{document}
	
\title{Neutrino pair annihilation above black-hole accretion disks in modified gravity}


\author{Gaetano Lambiase}
\affiliation{Dipartimento di Fisica ``E.R Caianiello'', Università degli Studi di Salerno, Via Giovanni Paolo II, 132 - 84084 Fisciano (SA), Italy}
\affiliation{Istituto Nazionale di Fisica Nucleare - Gruppo Collegato di Salerno - Sezione di Napoli, Via Giovanni Paolo II, 132 - 84084 Fisciano (SA), Italy}
\email{lambiase@sa.infn.it}

\author{Leonardo Mastrototaro}
\affiliation{Dipartimento di Fisica ``E.R Caianiello'', Università degli Studi di Salerno, Via Giovanni Paolo II, 132 - 84084 Fisciano (SA), Italy}
\affiliation{Istituto Nazionale di Fisica Nucleare - Gruppo Collegato di Salerno - Sezione di Napoli, Via Giovanni Paolo II, 132 - 84084 Fisciano (SA), Italy}
\email{lmastrototaro@unisa.it}
\correspondingauthor{Leonardo Mastrototaro}


\def\be{\begin{equation}}
\def\ee{\end{equation}}
\def\al{\alpha}
\def\bea{\begin{eqnarray}}
\def\eea{\end{eqnarray}}

\begin{abstract}
Using idealized models of the accretion disk, we investigate the effects induced by the modified theories of gravity on the annihilation of the neutrino pair annihilation into electron-positron pairs ($\nu{\bar \nu}\to e^-e^+$), occurring near the rotational axis. 
For the accretion disk, we have considered the models with temperature $T=constant$ and $T\propto r^{-1}$
%
%
In both cases, we find that the 
modified theories of gravity lead to an enhancement, up to more than one order of magnitude with respect to General Relativity, of the rate of the energy deposition rate of neutrino pair annihilation.
\end{abstract}

 \vskip -1.0 truecm

\section{Introduction}

Studies of GRB jets powered by the neutrino-antineutrino annihilation into electrons and positrons have quite a long history. The papers \cite{Co86,Co87,Goodman:1986we,1989Natur.340..126E,1993AcA....43..183J} were the first to analyze the energy deposition rate from the $\nu \bar{\nu}\rightarrow e^{+} e^{-}$ neutrino annihilation reaction, showing that this process can deposit an energy $\gtrsim 10^{51}$ erg above the neutrino-sphere of a type II supernova \cite{Goodman:1986we}. Salmonson \& Wilson in Ref.~\cite{Salmonson:1999es,Salmonson:2001tz} took into account strong gravitational field effects, which were included in a semi-analytic treatment. They showed that in a Schwarzschild geometry, the efficiency of the $\nu \bar{\nu}\rightarrow e^{+}e^{-}$ process is enhanced up to a factor of $30$ for collapsing neutron stars (NS) compared to the results in the Newtonian calculation.
A further study in this direction was considered in \cite{Asano:2000ib,Asano:2000dq}, where the authors studied the general relativistic effects on neutrino pair annihilation near the neutrinosphere and around a thin accretion disk for a Schwarzschild or Kerr metric. Here it was assumed that the accretion disk is isothermal.
Moreover, general relativity and rotation cause important effects in the spatial distributions of the energy deposition rate by $\nu $ and $\bar{\nu }$ annihilation \cite{Birkl:2006mu}.
In that paper, the deposition of energy and momentum in the vicinity of steady, axisymmetric accretion tori around stellar-mass black holes was investigated. The influence of general relativistic effects was investigated for different neutrinosphere properties. As compared to Newtonian calculations,
general relativistic effects increase the total annihilation rate measured by an observer at infinity by a factor of two when the neutrinosphere is a thin disk, recovering numerically the results of Ref.~\cite{Asano:2000dq}, but the factor is only $5/4$ for toroidal and spherical neutrinospheres. Thin disk models yield the highest energy deposition rates for neutrino-antineutrino annihilation and spherical neutrinospheres the lowest ones, independently of whether general relativistic effects are included or not. It is worth mentioning that the neutrino annihilation luminosity from the disk has been calculated also by other authors assuming various conditions (see also \cite{Mallick:2008iv,Bhattacharyya:2009nm,Chan:2009mw,Kovacs:2009dv,Kovacs:2010zp,doi:10.1063/1.3155863,Harikae_2010}). Moreover, various time-dependent models of black-hole accretion disks as remnants of neutron-star mergers or collapsar engines exist, of which most estimate the pair-annihilation rates in a post-processing manner (e.g.~\cite{Harikae_2010,1999A&A...344..573R,Popham_1999,Di_Matteo_2002}), while a few recent ones include pair-annihilation and its dynamical impact self-consistently during the evolution~\cite{Fujibayashi_2017,Just:2015dba,PhysRevD.98.063007,2020ApJ...902L..27F}. These works indicate that. that neutrino-pair annihilation in ordinary GR models seems to be not efficient enough to power GRBs and the Blandford-Znajek process is currently considered a more promising mechanism for launching jets. Finally, the neutrino-antineutrino annihilation into electron-positron pairs emitted only from the surface of a neutron star assuming that the gravitational background is described by modified theories of gravity has been investigated in \cite{Lambiase:2020iul,Lambiase:2020pkc}.

The aim of this paper is to study the general relativistic effects (induced by modified gravity) on the energy deposition of neutrinos emitted from a thin accretion disk (we follow the seminal papers \cite{Asano:2000dq,Asano:2000ib}). 
Modified theories of gravity have been invoked to solve still open questions that make GR incomplete, that arise at short distances and small time scales (black hole and cosmological singularities, respectively), and at large distance scales (the rotational curve of the galaxies and the observed accelerated phase of the present Universe \cite{riess}). 
Deviations from the GR (hence from the Hilbert-Einstein action) allow to solve these issues \cite{Capozziello:2011et,cosmo8,cosmo9,odi,Amendola:2015ksp,DeFelice:2010aj,Clifton:2011jh,Dainotti:2022bzg}. In the last years, several {\it alternative} or {\it modified} theories of gravity have been proposed, which try to answer the above-mentioned issues of GR and the Cosmological Standard Model. One of the consequences, which we are interested in, is that the metric tensor $g_{\mu\nu}$ describing the gravitational field generated by a massive source gets modified with respect to the Schwarzschild metric. The standard solutions obtained in GR are recovered in the limit in which the parameters characterising some specific theory of gravity beyond GR are set equal to zero.

In our analysis, we model the disk temperature profile with $T=constant$ or $T=2R_{ph}/r$, with $R_{ph}$ the photosphere radius, being the simplest possible temperature models~\cite{Asano:2000dq,Asano:2000ib,10.1143/PTPS.136.235}. 

The plan of our work is as follows. In Sec.~\ref{Formulation} we present the model used for computing the energy deposition from the thin disk. In Sec.~\ref{Results} we characterise the effects of the theories beyond General Relativity on the energy deposition rate of neutrino pair annihilation near the rotational axis of the gravitational source. Finally, in Sec.~\ref{Conclusion} we summarise our results.

\section{Formulation}
\label{Formulation}

Let us consider a BH with a thin accretion disk around it that emits neutrinos (see \cite{Asano:2000dq}). We will confine ourselves to the case of an idealised, semi-analytical, stationary state model, which is independent of details regarding the disk formation. The disk is described by an inner and outer edge, with corresponding radii defined by $R_{\mathrm{in}}$ and $R_{\mathrm{out}}$, respectively. Self-gravitational effects are neglected. The generic metric with a spherical symmetry is given by
\begin{equation}
    g_{\mu\nu}=\left(g_{00},g_{11},-r^2,-r^2\sin^2\theta\right) \,.
\end{equation}
The Hamiltonian of a test particle moving in this geometry is defined as
\begin{equation}
2\mathcal{H}=-E\dot{t}+L\dot{\phi}+g_{11}\dot{r}^2=\delta_1 \,,
\end{equation}
where $\delta_1=0$ for null geodesics and $\delta_1 =1$ for massive particles, while $E$ and $L$ are,  respectively,  the energy and angular momentum of the test particles moving around the rotational axis of the BH. With the above definitions, one obtains the non-vanishing components of the 4-velocity \citep{Prasanna:2001ie}
\begin{align}
U^{3}&=\dot{\phi}=-\frac{L}{r^2} \,\ ; \\
U^0&=\dot{t}=-\frac{E}{g_{00}} \,\ ; \\
\dot{r}^2&=\frac{E\dot{t}-L\dot{\phi}}{g_{11}} \,\ ,
\label{dr/dt}
\end{align}

We are interested in the energy deposition rate near the rotational axis at $\theta=0^{\circ}$. We use the value $\theta=0^{\circ}$ for evaluating the energy emitted in a half cone of $\Delta \theta\sim10^{\circ}$. The scalar product of the moment of a neutrino and antineutrino at $\theta=0^{\circ}$ can be cast in the form
\begin{equation}
    p_{\nu}\cdot p_{\bar{\nu}} = E_{\nu}E_{\bar{\nu}}\left[1-\sin\theta_{\nu}\sin\theta_{\bar{\nu}}\cos\left(\phi_{\nu}-\phi_{\bar{\nu}}\right)-\cos\theta_{\nu}\cos\theta_{\bar{\nu}}\right] \,\ ,
\end{equation}
where $E_{\nu}=E_{0\nu}/\sqrt{g_{00}}$, $E_{0\nu}$ is the observed energy of the neutrino at infinity and
\begin{equation}
    \sin\theta_{\nu}=\frac{\rho_{\nu}}{r}\sqrt{g_{00}(r)} \,\ ,
    \label{sen}
\end{equation}
with $\rho_{\nu}=L_{\nu}/E_{0\nu}$. It is important to notice that, for geometrical reasons, there exist a minimal and maximal value $\theta_m$ and $\theta_M$ for a neutrino coming from $R_{in}=2R_{\mathrm{ph}}$ and $R_{out}=30M$ respectively, with $R_{\mathrm{ph}}$ the photosphere radius. Moreover, it can be shown that the following relation holds~\cite{Asano:2000dq}
\begin{equation}
    \rho_{\nu}=\frac{r_0}{\sqrt{g_{00}(r_0)}} \,\ ,
    \label{rho}
\end{equation}
with $r_0$ the nearest position between the particle and the centre before arriving at $\theta=0$. Finally, we also need the equation of the trajectory \cite{Asano:2000dq}
\begin{equation}
    \frac{\pi}{2}=\int_C\frac{dr'}{r'\sqrt{(r'/\rho_{\nu})^2-g_{00}(r')}} \,.
    \label{trajectory equation}
\end{equation}
In this relation one takes into account that the neutrinos are emitted from the position $(R,\pi/2)$, with $R\in [R_{in},R_{out}]$, and arrive at $(r,0)$. The energy deposition rate of neutrino pair annihilation is therefore given by \cite{Asano:2000dq}:
\begin{equation}
    \frac{dE_0(r)}{dtdV}=\frac{21\pi^4}{4}\zeta(5)KG_F^2k^9T^9_{\mathrm{eff}}(2R_{ph})F(r) \,\ ,
    \label{trajectory}
\end{equation}
where $G_F$ is the Fermi constant, $k$ is the Boltzmann constant, $T_{\mathrm{eff}}(2R_{\mathrm{ph}})$ is the effective temperature at radius $2R_{ph}$ (the temperature observed in the comoving frame),
\begin{equation}
    K=\frac{1\pm 4\sin^2\omega_W+8\sin^4\theta_W}{6\pi} \,\ ,
\end{equation}
with the $+$ sign for $\nu_e$ and the $-$ sign for $\nu_{\mu/\tau}$, $\sin^2\theta_W=0.23$ ($\theta_W$ is the Weinberg angle), and finally
\begin{equation}
\begin{split}
    &F(r)=\frac{2\pi^2}{T^9_{\mathrm{eff}}(2R_{ph})}\frac{1}{g_{00}(r)^4}\Bigg(2\int_{\theta_m}^{\theta_M}d\theta_{\nu}T_0^5(\theta_{\nu})\sin\theta_{\nu}\int_{\theta_m}^{\theta_M}d\theta_{\bar{\nu}}T_0^4(\theta_{\bar{\nu}})\sin\theta_{\bar{\nu}}+\\&
    +\int_{\theta_m}^{\theta_M}d\theta_{\nu}T_0^5(\theta_{\nu})\sin^3\theta_{\nu}\int_{\theta_m}^{\theta_M}d\theta_{\bar{\nu}}T_0^4(\theta_{\bar{\nu}})\sin^3\theta_{\bar{\nu}}+\\ 
    &+2\int_{\theta_m}^{\theta_M}d\theta_{\nu}T_0^5(\theta_{\nu})\cos^2\theta_{\nu}\sin\theta_{\nu}\int_{\theta_m}^{\theta_M}d\theta_{\bar{\nu}}T_0^4(\theta_{\bar{\nu}})\cos^2\theta_{\bar{\nu}}\sin\theta_{\bar{\nu}}- \\&
    -4\int_{\theta_m}^{\theta_M}d\theta_{\nu}T_0^5(\theta_{\nu})\cos\theta_{\nu}\sin\theta_{\nu}\int_{\theta_m}^{\theta_M}d\theta_{\bar{\nu}}T_0^4(\theta_{\bar{\nu}})\cos\theta_{\bar{\nu}}\sin\theta_{\bar{\nu}}\Bigg) \,\ ,
    \end{split}
    \label{F(r)}
\end{equation}
with $T_0$ the temperature observed at infinity
\begin{align}
    T_0(R)&=\frac{T_{\mathrm{eff}}(R)}{\gamma}\sqrt{g_{00}(R)} \label{Teff} \,\ , \\
    \gamma&=\frac{1}{\sqrt{1-v^2/c^2}} \,\ , \\
    \frac{v^2}{c^2}&=\frac{g_{33}}{g_{00}}\frac{g_{00,r}}{2r} \label{v^2} \,\ ,
\end{align}
where $T_{\mathrm{eff}}$ is the effective temperature measured by a local observer and all the quantities are evaluated at $\theta=\pi/2$. In the treatment we will ignore the reabsorption of the deposited energy by the black hole and we will consider the case of isothermal disk ($T_{\mathrm{eff}}\sim\mathcal{O}(10)~\rm{MeV}$) and the case of a simple gradient temperature~\cite{Asano:2000dq} ($T_{\mathrm{eff}}\propto r^{-1}$). In particular, we take (for details, see \cite{10.1143/PTPS.136.235})
\begin{equation}\label{tdippr}
T_{\mathrm{eff}}(r)\propto \frac{2R_{ph}}{r} \,\ . 
\end{equation}
The assumptions of the value of the temperature and the shape of the gradient model are compatible with recent results of neutrino-cooled accretion disk models (e.g. \cite{2015ApJ...805...37L,Liu:2007bca,2007ApJ...662.1156K}). 
It is usually expected that the effective maximum temperature $T_{\mathrm{eff}}$ is $\sim\mathcal{O}(10~\mathrm{MeV})$. This comes, from a phenomenology point of view,  from the disk neutrino luminosity, which is strictly dependent on $T$. The latter, therefore, can not be substantially different among various models. Moreover, since we are not performing numerical simulations, we assume $T_{\mathrm{eff}}\sim\mathcal{O}(10~\mathrm{MeV})$ to compare the effects of different gravitational models in the same conditions.\footnote{The maximum value of $T_{\mathrm{eff}}$, in a classical dynamic treatment, is  $T\propto \dot{M}/R_{\mathrm{in}}^3$, with $\dot{M}$ the accretion rate (the material that falls inside the BH). For a disk with a constant density, the solution of the Navier-Stokes equations gives $\dot{M}\propto u_r \rho\sqrt{g_{11}(R_{\mathrm{in}})} R_{\mathrm{in}}$~\cite{Narayan:1998ft}, where $u_r$ is the radial velocity of the material and $\rho\sqrt{g_{11}(R_{\mathrm{in}})}$ takes into account the surface density. Even in this simple treatment, the maximum temperature accounts for two undetermined parameters that might have an important role and can be determined only by a numerical disk simulation. However, considering $u_r$ and $\rho$ constant, the values of the maximum temperature are almost equal in each model.} We finally remark that, despite all these theoretical assumptions, only a simulation of the disk from NS-NS merging with assigned geometry would give the exact profile of the temperature.

\section{Applications to modified theories of gravity}
\label{Results}
In this section, we calculate the emitted energy with the procedure shown in Sec.~\ref{Formulation} for different modified geometries, that is for modified theories of gravity with respect to GR. The enhancement of the energy deposition rate discussed in this Section depends on the modification, with respect to GR, of the neutrino trajectory and of the temperature $T_{0}(R)$ (see Eqs.~(\ref{trajectory equation}) and (\ref{Teff})). 

We mainly focus on those models of modified theories of gravity that present an analytic solution for the spherical symmetric geometry and allow for an enhancement of the 
energy deposition rate compared to GR. In particular, we consider the black hole surrounded by Quintessence and Non-linear electrodynamics, Charged Galileon and Kerr-Sen geometries, which are characterised by the fact that the modifications to GR are induced by dark energy effects, different couplings to the Standard Model interactions, additional fields or modification arising from string theory respectively (we have also investigated other models, such as Eddington-inspired Born-Infeld black hole solution~\cite{Mukherjee:2017fqz}:, Brans Dicke gravity~\cite{Brans:1961sx} and Higher derivative gravity analytical solution~\cite{Kokkotas:2017zwt}, but they do not produce a relevant enhancement of the energy deposition rate compared to GR). Moreover, the recent GW observations
allow for constraining modified theory of gravity \cite{PhysRevLett.116.221101,Baker:2017hug,Creminelli:2017sry,Sakstein:2017xjx,Carson:2019rda,Gnocchi:2019jzp,LIGOScientific:2019fpa}, As shown in \cite{Ezquiaga:2017ekz}, only derivatives of orders greater to one are constrained by the GW detection because they give rise to anomalous GWs speed. The latter condition does not occur in the models considered in our paper\footnote{The propagation of GWs in theories of gravity beyond GR is, in general, quite involved since the additional fields (scalar or electromagnetic in the cases analysed in this paper), hence the additional degrees of freedom, might modify the background over which GWs propagate and their perturbations might mix with the metric ones. The models here discussed do not affect the GW propagation (see for example \cite{Ezquiaga:2018btd}).}.
We also mention that the black hole shadow of these models has been investigated in \cite{Zeng:2020vsj} for black holes surrounding by quintessence, in \cite{Stuchlik2019,Okyay:2021nnh,Stuchlik:2019uvf} for non-linear electrodynamics black holes, in \cite{Tretyakova:2016ale} for Horndeski/Galileon theory, and finally in \cite{Lima:2021las,Xavier:2020egv,Wu:2021pgf} for Kerr-Sen model\footnote{Results show that all these models are compatible with the recent observations of the Event Horizon Telescope that has been able to capture, by using very long baseline interferometry, the images of the black hole shadow of a super-massive M87
blackhole \cite{EventHorizonTelescope:2019dse,EventHorizonTelescope:2019ths}. It must be pointed out, however, that the origin and nature of GRBs is still an open issue in astrophysics.}.

\subsection{Quintessence}
We first consider the spacetime around a black hole surrounding by quintessence. 
A static spherically-symmetric black hole surrounded by quintessence in $d$-dimensions is described by \cite{Chen:2008ra}
\begin{eqnarray}\label{AGae1}
ds^2=e^{\nu(r)}dt^2-e^{\lambda(r)}dr^2-r^2d\theta^2_1-r^2\sin^2{\theta_1}d\theta^2_2-\cdots-r^2\sin^2{\theta_1}
\cdots\sin^2{\theta_{d-3}}d\theta^2_{d-2}\,,
\end{eqnarray}
where $d$ are the dimensions of our space-time. For a static spherically symmetric configuration, the quintessence energy density tensor is \cite{Kiselev:2002dx}
\begin{equation}\label{Tcomps}
T^{\;0}_0 = A(r), \quad T^{\;j}_0=0,\quad T^{\;j}_i=C(r)r_i r^j+B(r)\delta^{\;j}_i.
\end{equation}
Considering the average over the angles of isotropic state $\langle r_ir^j\rangle =\frac{1}{d-1}r_k r^k \delta^{\;j}_i$, it is possible to obtain
\begin{equation}
\langle T^{\;j}_i\rangle = D(r)\delta^{\;j}_i,\quad D(r)=-\frac{1}{d-1}C(r)r^2+B(r).
\end{equation}
Furthermore the quintessence is characterised by the condition $p =-\omega_q \rho$ which implies $D(r)=-\omega_q A(r)$. Following the derivation in Ref. \cite{Kiselev:2002dx,Chen:2008ra}  one considers $C(r) \propto B(r)$, so that the exact solutions are possible. The constant coefficient between $C(r)$  and $B(r)$  is defined by the additivity and linearity condition\footnote{Such a condition allows to obtain the correct limits for the well-known cases of charged black holes ($w_q = 1/3$), dust matter ($w_q=0$), and quintessence ($w_q = -1$)  \cite{Kiselev:2002dx}.}. It then follows that 
the metric of the spherically symmetric black hole surrounded by quintessence in $d=4$ dimensions is given by
\begin{eqnarray}\label{A3}
ds^2=\bigg[1-\frac{2M}{r}-\frac{c}{r^{3\omega_q+1}}\bigg]dt^2
-\frac{dr^2}{\displaystyle{1-\frac{2M}{r}-\frac{c}{r^{3\omega_q+1}}}} -r^2d\Omega,
\end{eqnarray}
with $c$ a positive constant and $-1<\omega_q<1/3$.
\subsubsection{Isothermal model}
\begin{figure*}
\gridline{\fig{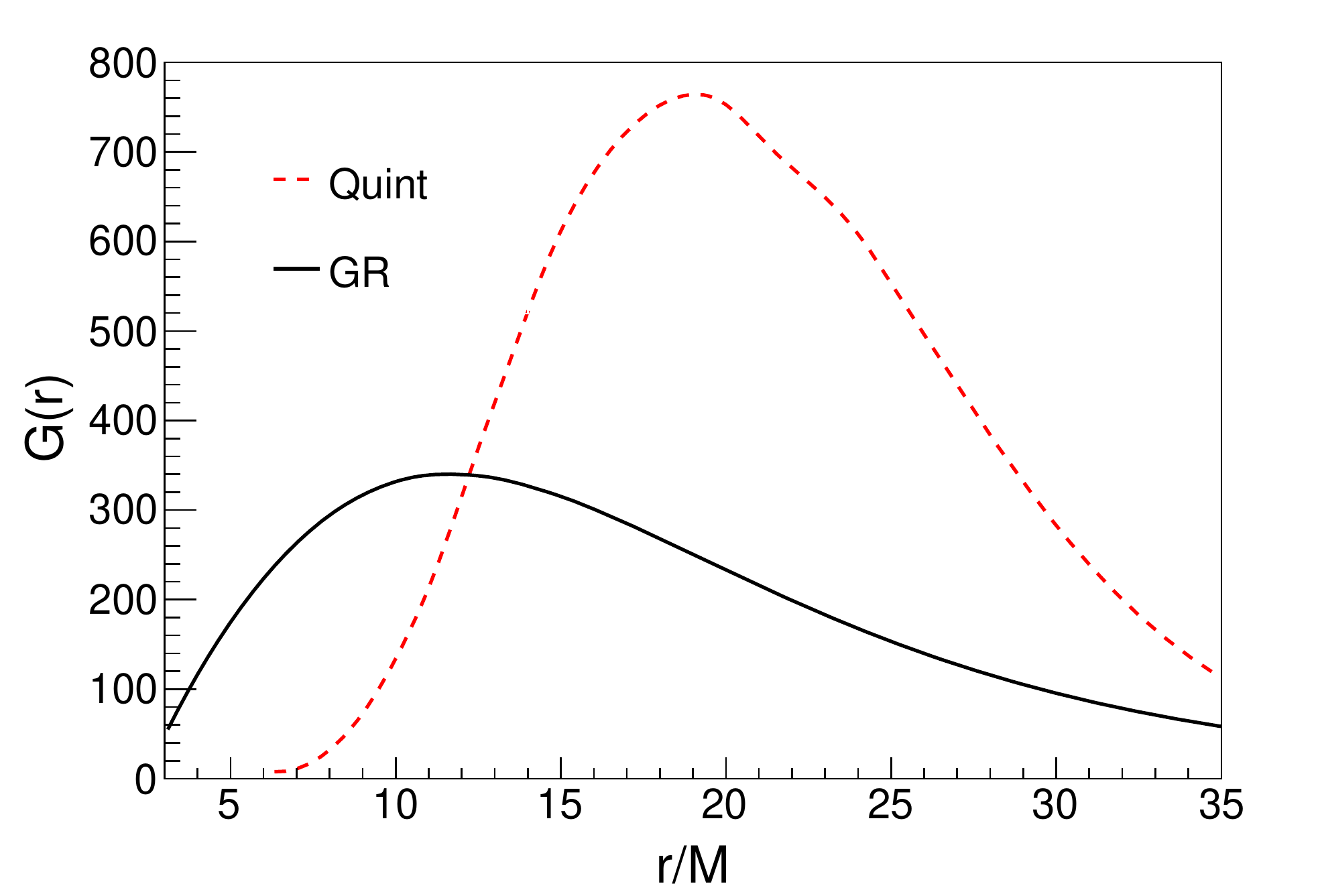}{0.49\textwidth}{(a)}
          \fig{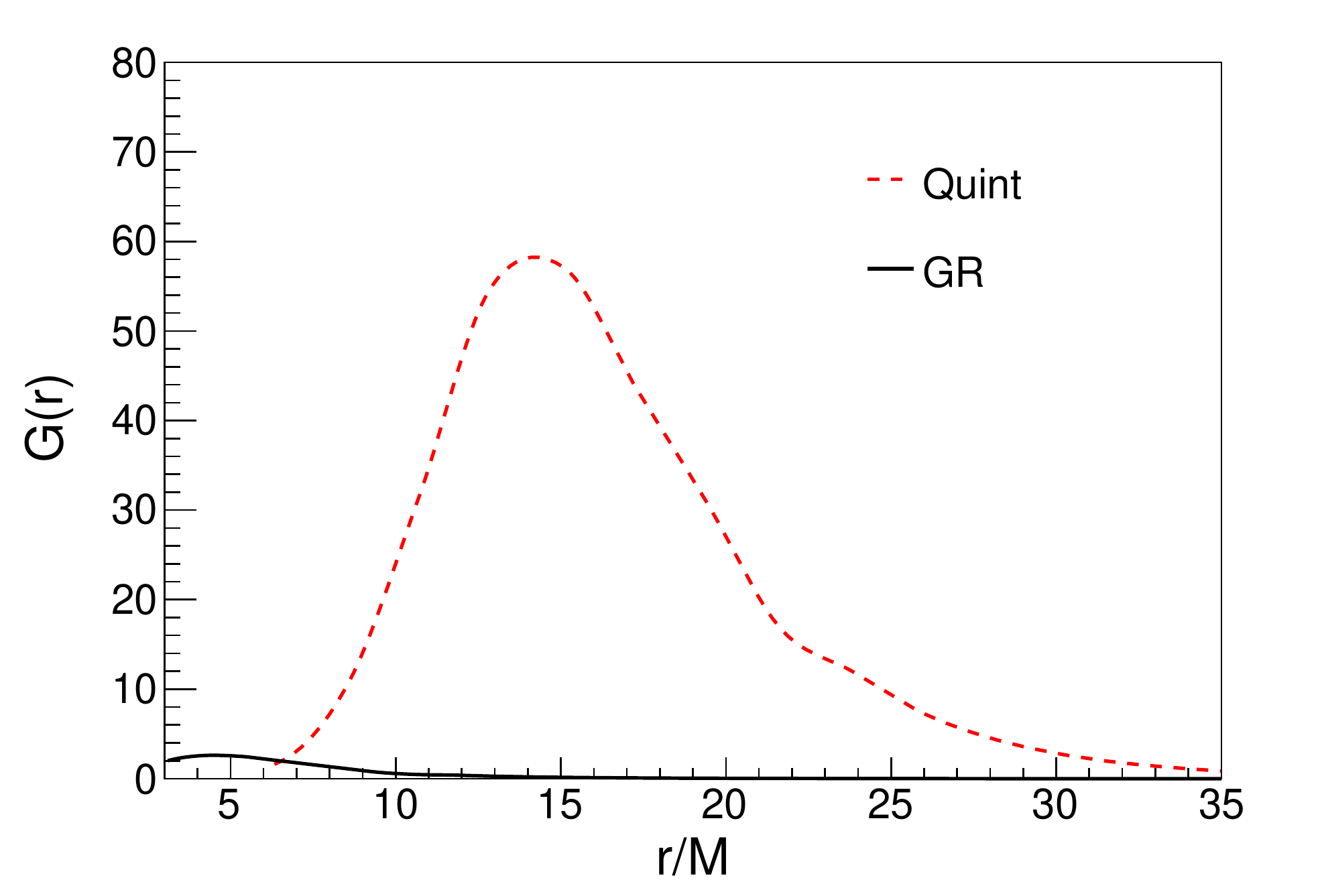}{0.49\textwidth}{(b)}
          }
\caption{Plot of $G(r)$ against $r/M$ for the isothermal disk (a) and temperature gradient model (b). The solid black line is for GR while the dashed red line is for the quintessence metric with $c=0.4$ and $\omega=-0.4$.}
\label{fig:GR}
\end{figure*}

Using Eq.~(\ref{F(r)}), it is possible to obtain the results in Fig.~\ref{fig:GR}~(a), where we have plotted the function $G(r)$ defined as:
\begin{equation}\label{G(r)Gae}
    G(r)=F(r)\frac{r^2}{4M^2} \,\ .
\end{equation}
The function $G(r)$ is essential to evaluate the energy deposition rate (EDR) and, therefore, the energy viable for a GRB explosion. We estimate the EDR in the infinitesimal angle $d\theta$ taking into consideration a characteristic angle of $10^{\circ}$ and temperature of $10~\mathrm{MeV}$~\cite{Asano:2000dq}:
\begin{equation}
    \frac{dE_0}{dt}\simeq 4.41\times 10^{48}\left(\frac{\Delta \theta}{10^\circ}\right)^2\left(\frac{kT_\mathrm{eff}(R_{\mathrm{in}})}{10~\mathrm{MeV}}\right)^9\left(\frac{2M}{10~\mathrm{km}}\right)\int_{R_{\mathrm{in}}}^{R_{\mathrm{out}}}\frac{G(r)}{2M}dr~\mathrm{erg~s^{-1}} \,\ .
    \label{value}
\end{equation}

In Fig.~\ref{fig:GR}~(a) we show the behaviour of $G(r)$ in GR (continuous black curve) and Quintessence metric (dashed red curve) with $c=0.4$ and $\omega_q=-0.4$, to show the order of magnitude of enhancement that it is possible to obtain with this metric. Initially, $G(r)$ increases with the distance, reaching
a maximum value, and then, due to the interplay between temperature and red-shift effects, decreases with distance; this shape is common between all the metrics, as we will show in the following subsections.
The energy viable for a GRB emission is given by
\begin{align}
\frac{dE_0^{GR}}{dt}&\simeq 1.5\times 10^{52}~\mathrm{erg~s^{-1}} \,\ , \\
\frac{dE_0^{Quint}}{dt}&\simeq 3.0\times 10^{52}~\mathrm{erg~s^{-1}} \,\ .
\label{EDRQuint}
\end{align}
For late discussion, it turns out useful to define the enhancement factor:
\begin{equation}
    \Pi_{(T=\mathrm{cost})}^{Quint} \equiv \frac{dE_0^{Quint}/dt}{dE_0^{GR}/dt}\sim 2 \,\ .
\end{equation}
Notice that the small value of $ \Pi_{(T=\mathrm{cost})}^{Quint}$ is caused by $R_{\mathrm{in}}^{\mathrm{GR}}\gg R_{\mathrm{in}}^{\mathrm{Quint}}$. Indeed, $T_{\mathrm{eff}}=constant$ means that even neutrinos emitted from $R\sim R_{out}$ significantly contribute to the EDR. This is equivalent to affirm that the EDR depends on the disk extension. The enhancement would be larger if we would set for both metrics $R_{\mathrm{in}}=R_{\mathrm{in}}^{\mathrm{Quint}}$. 

\subsubsection{Temperature gradient model}
\label{Quintessence - Temperature gradient}
Using Eq.~(\ref{F(r)}), considering that the temperature varies along $r$ and thus along $\theta_{\nu(\bar{\nu})}$,it is possible to obtain the results in of Fig.~\ref{fig:GR}~(b), where we have plotted the function $G(r)$ for GR (continuous black curve) and quintessence metric (dashed red curve), with $c=0.4$ and $\omega_q=-0.4$. One obtains that the energy viable for a GRB emission is
\begin{align}
\frac{dE_0^{GR}}{dt}&\simeq 4.5\times 10^{49}~\mathrm{erg~s^{-1}} \,\ , \\
\frac{dE_0^{Quint}}{dt}&\simeq 1.3\times 10^{51}~\mathrm{erg~s^{-1}} \,\ , 
\label{EDRQuint1}
\end{align}
and as for constant temperature, we define the enhancement factor as
\begin{equation}
\Pi_{(T\propto 1/r)}^{Quint} =\frac{dE^{Quint}_0/dt}{dE^{GR}_0/dt}\sim 30 \,\ .
\end{equation}
Some comments are in order. Firstly, as one can see, even in the case of the temperature gradient  $T\propto R_{in}/r$, the quintessence model may provide the energy of the emitted GRB, differently from what happens in GR. Moreover, from Fig.~\ref{fig:GR} one can see that $\Pi_{(T\propto 1/r)}^{Quint}>\Pi_{(T=cost)}^{Quint}$. This peculiar effect is due to the finite size of the disk, as already explained in the previous sub-section. The enhancement in the $T=constant$ model is attenuated because it depends on the disk extension. As a consequence, the effect of the modification of the trajectory of neutrinos is compensated by the smaller disk extension, compared to GR, as it arises from Fig. \ref{fig:GR} (a). In the $T\propto R_{in}/r$ model we can appreciate the enhancement without the counter-effect of the disk extension because, due to the $T$ reduction over distances, the neutrinos from the disk contribute to the energy deposition until a radius $R\ll R_{\mathrm{out}}$. Finally, the EDR in the gradient model, Eq. (\ref{EDRQuint1}), is reduced as compared to the isothermal model, Eq. (\ref{EDRQuint}), as follows from the comparison between Fig. \ref{fig:GR} (a) and (b). This behaviour is common in the model with $T\propto r^{-1}$ since the energy deposition rate depends on the value of the temperature, as shown in Eq. (\ref{F(r)}).

\subsection{Charged Galileon}
We consider now the charged Galileon black holes, which are a subclass of Horndeski theories. This BH solution inherits an additional gauge field that is nonminimally coupled to the scalar sector. The action takes the form of~\citep{Mukherjee:2017fqz,Babichev:2015rva}
\begin{equation}
\begin{split}
S=&\frac{1}{16\pi}\int d^4x\sqrt{-g}\Bigg[R-\frac{1}{4}F_{\mu\nu}F^{\mu\nu}+\beta G^{\mu\nu}\nabla_{\mu}\psi\nabla_{\nu}\psi-\eta\partial_{\mu}\psi\partial^{\mu}\psi-\\&-\frac{\gamma}{2}\left(F_{\mu\sigma}F_{\nu}^{\sigma}-\frac{1}{4}g_{\mu\nu}F^{\alpha\beta}F_{\alpha\beta}\right)\nabla^{\mu}\psi\nabla^{\nu}\psi\Bigg] \,\ ,
\end{split}
\end{equation}
where $\beta$ and $\eta$ are coupling constants with $\beta\neq 0$, $\psi$ is the Galileon field, $-1/4F_{\mu\nu}F^{\mu\nu}$ is the canonical term for the gauge field and $G^{\mu\nu}$ is the Einstein tensor.  Imposing spherical condition for the defined action, one obtains the following metric~\citep{Mukherjee:2017fqz,Babichev:2015rva}
\begin{equation}\label{Galmetric}
    g_{00}=g_{11}^{-1}=1-\frac{2M}{r}-\frac{\Lambda r^2}{3}+\frac{M^2q}{r^2} \,\ ,
\end{equation}
where $\Lambda=\eta/2\beta$ is related to dark energy and $q$ is the charge of the BH.
\subsubsection{Isothermal model}
\begin{figure*}
\gridline{\fig{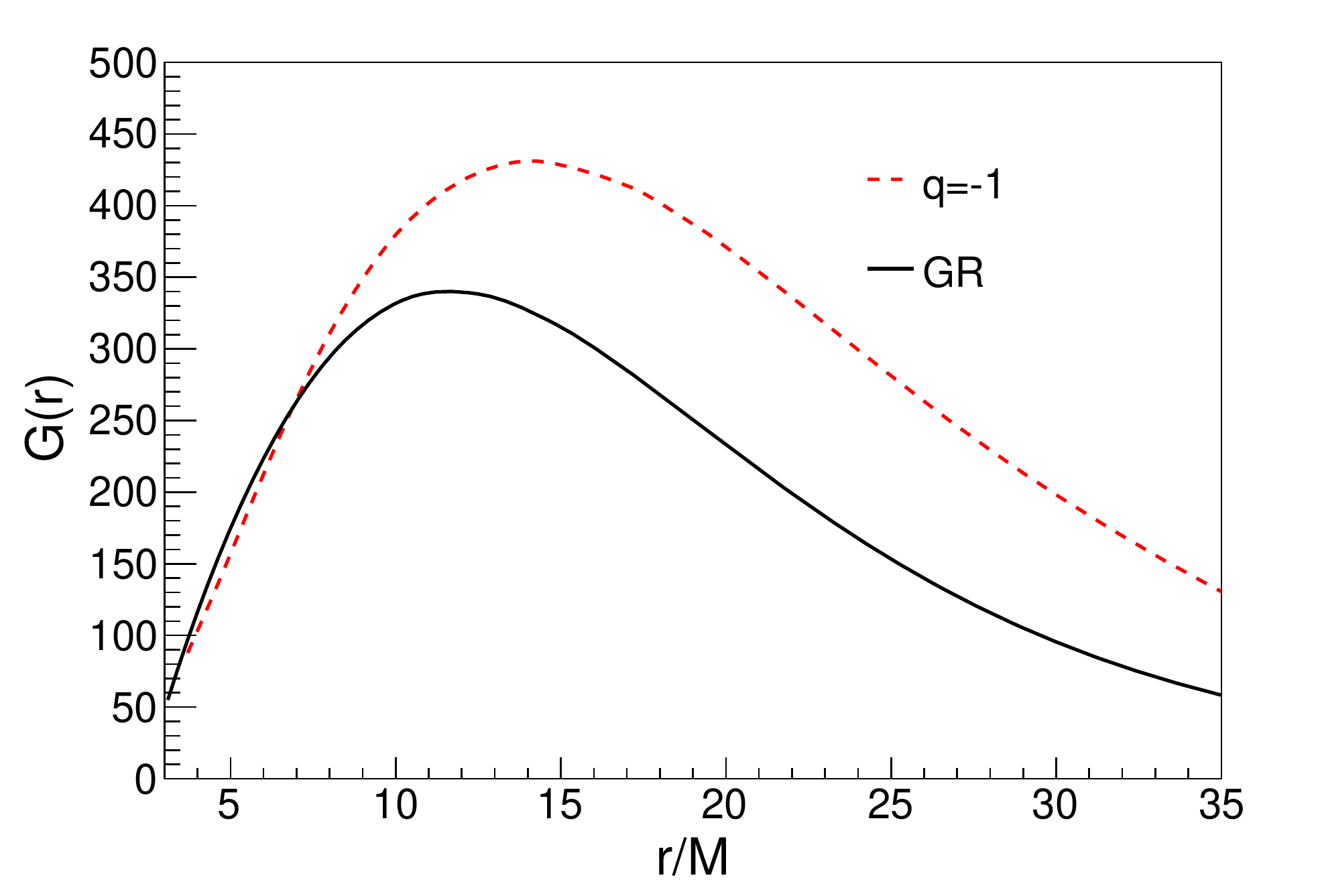}{0.49\textwidth}{(a)}
          \fig{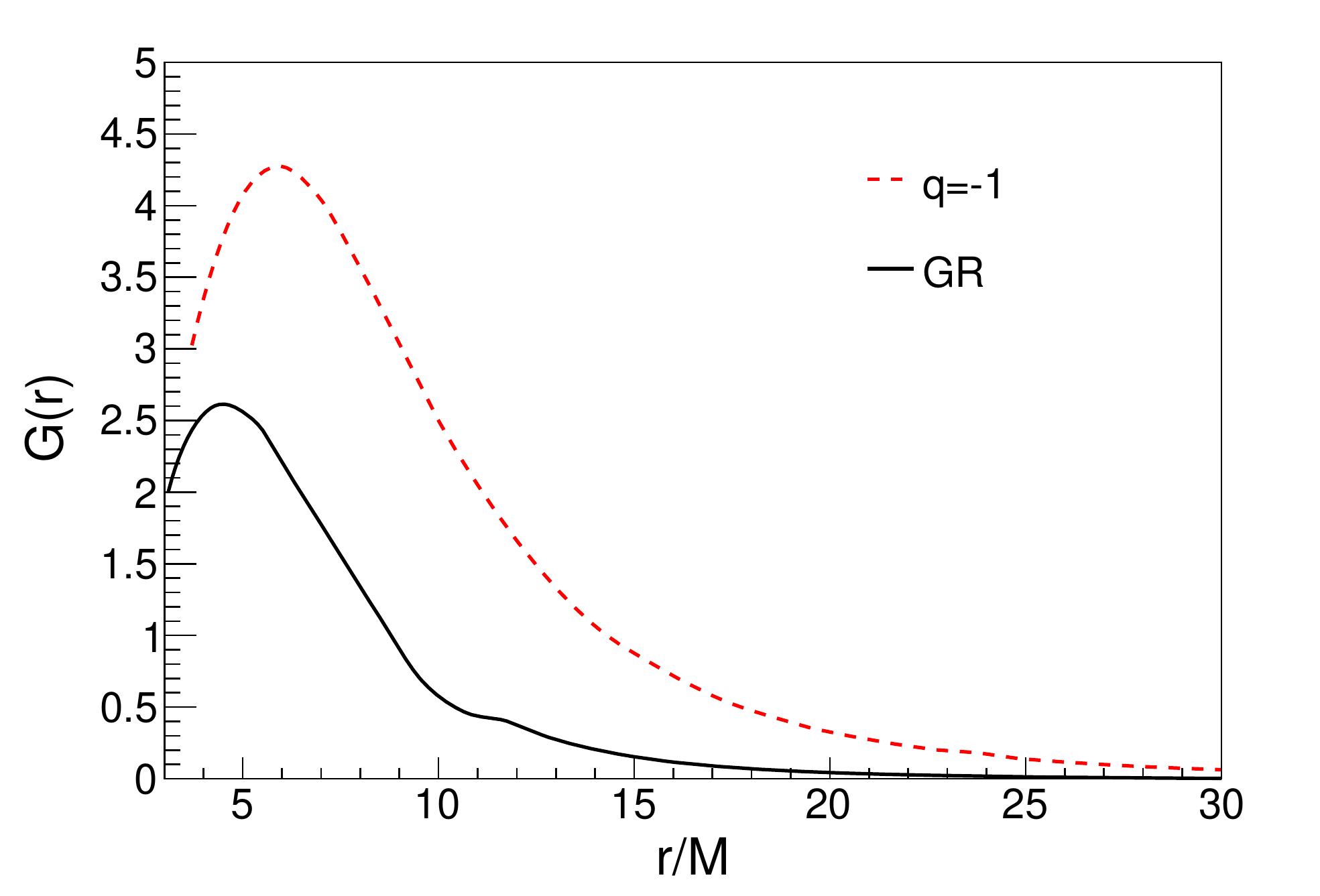}{0.49\textwidth}{(b)}
          }
\caption{Plot of $G(r)$ against $r/M$ for the isothermal disk (a) and temperature gradient model (b). The solid black line is for GR while the dashed red line is for the charged galileon metric with $q=-1$ and $\Lambda=10^{-3}$.}
\label{fig:q-1}
\end{figure*}

 For the model with $T=constant$, the function $G(r)$ entering  in (\ref{G(r)Gae}) is plotted in Fig.~\ref{fig:q-1}~(a) for GR (continuos black curve) and charged galileon metric (dashed red curve) with $\Lambda=10^{-3}M^{-2}$ and $q=-1$, showing the order of magnitude of the enhancement that it is possible to obtain with this model. The EDR is only slightly superior to that of GR:
\begin{align}
    \frac{dE^{CG}}{dt}&\simeq 2.2\times 10^{52} \mathrm{erg~s^{-1}} \,\ , \\
    \Pi_{(T=\mathrm{cost})}^{CG}&\sim 1.4 \,\ .
\end{align}
Furthermore, we want to emphasise that the contribute of $\Lambda$ to the enhancement is negligible, similar to that shown in Ref.~\cite{Lambiase:2020iul}.

\subsubsection{Temperature gradient model}
As done in Sec.~\ref{Quintessence - Temperature gradient}, using Eq.~(\ref{F(r)}) and considering $T\propto 2R_{\mathrm{ph}}/r$, it is possible to obtain the function $G(r)$ in Fig.~\ref{fig:q-1}~(b) for GR (black continuous line) and the charged galileon metric (red dashed line), with $\Lambda=10^{-3}M^{-2}$ and $q=-1$.

As in the case of the Quintessence, the enhancement with respect to GR is still present and we obtain
\begin{align}
    \frac{dE^{CG}}{dt}&\simeq 1.1\times 10^{50} \mathrm{erg~s^{-1}} \,\ , \\
    \Pi_{(T\propto 1/r)}^{CG}&\sim 2.5 \,\ .
\end{align}

\vspace{0.2in}

In the metric (\ref{Galmetric}) we do not have peculiar differences between the two models $T=constant$ and $T\propto 1/r)$. Indeed, the enhancement factors $\Pi_{(T=\text{const})}^{CG}$ and $\Pi_{(T\propto 1/r)}^{CG}$ are of the order $\sim 2$ in both cases.
\subsection{Nonlinear electrodynamics}

This theory arises from a coupling of the Einstein gravity can be coupled with nonlinear
electromagnetic field of the type~\cite{Fan:2016hvf} (see also \cite{,Bronnikov:2017tnz,PhysRevD.96.128502,Nojiri:2017kex,PhysRevD.98.028501,Toshmatov:2018ell})
\begin{equation}
    I=\frac{1}{16\pi}\int d^4x\sqrt{-g}(R-\mathcal{L}(\mathcal{F})) \,\ ,
\end{equation}
where $\mathcal{F}=F_{\mu\nu}F^{\mu\nu}$ and 
\begin{equation}
    \mathcal{L}=\frac{4\mu}{\alpha}\frac{(\alpha\mathcal{F})^{5/4}}{(1+\sqrt{\alpha}\mathcal{F})^{1+\mu/2}} \,\ ,
\end{equation}
Imposing the spherical symmetry, one obtains the following metric \cite{Fan:2016hvf}
\begin{equation}
    g_{00}=g_{11}^{-1}=1-\frac{2M}{r}-\frac{2\alpha^{-1}q^3r^{\mu-1}}{(r^2+q^2)^{\mu/2}} \,\ ,
    \label{new_metric}
\end{equation}
where $q$ is related to the charge of the black hole and $\alpha$ and $\mu$ come from the nonlinear coupling. 

\subsubsection{Isothermal model}

\begin{figure*}
\gridline{\fig{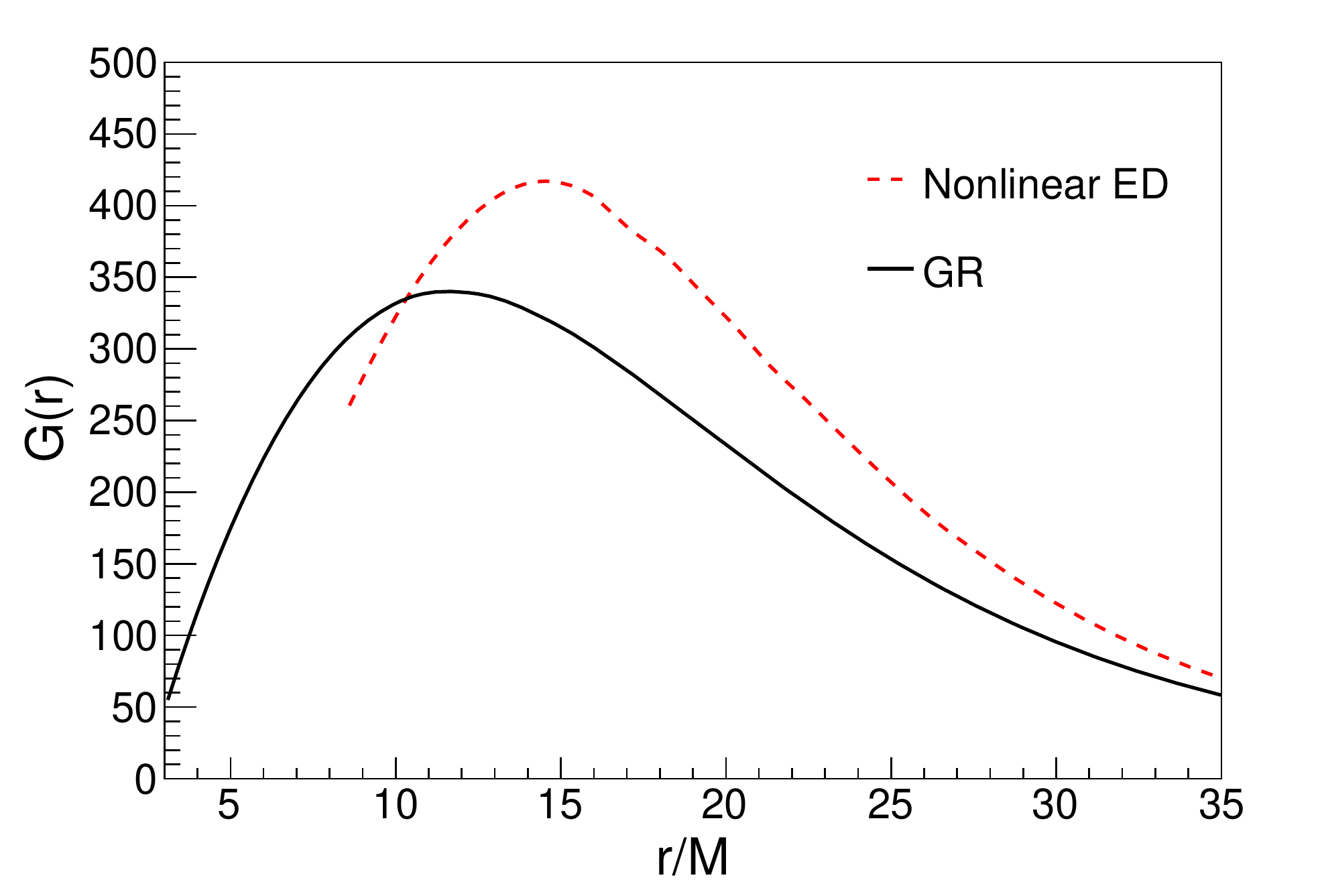}{0.49\textwidth}{(a)}
          \fig{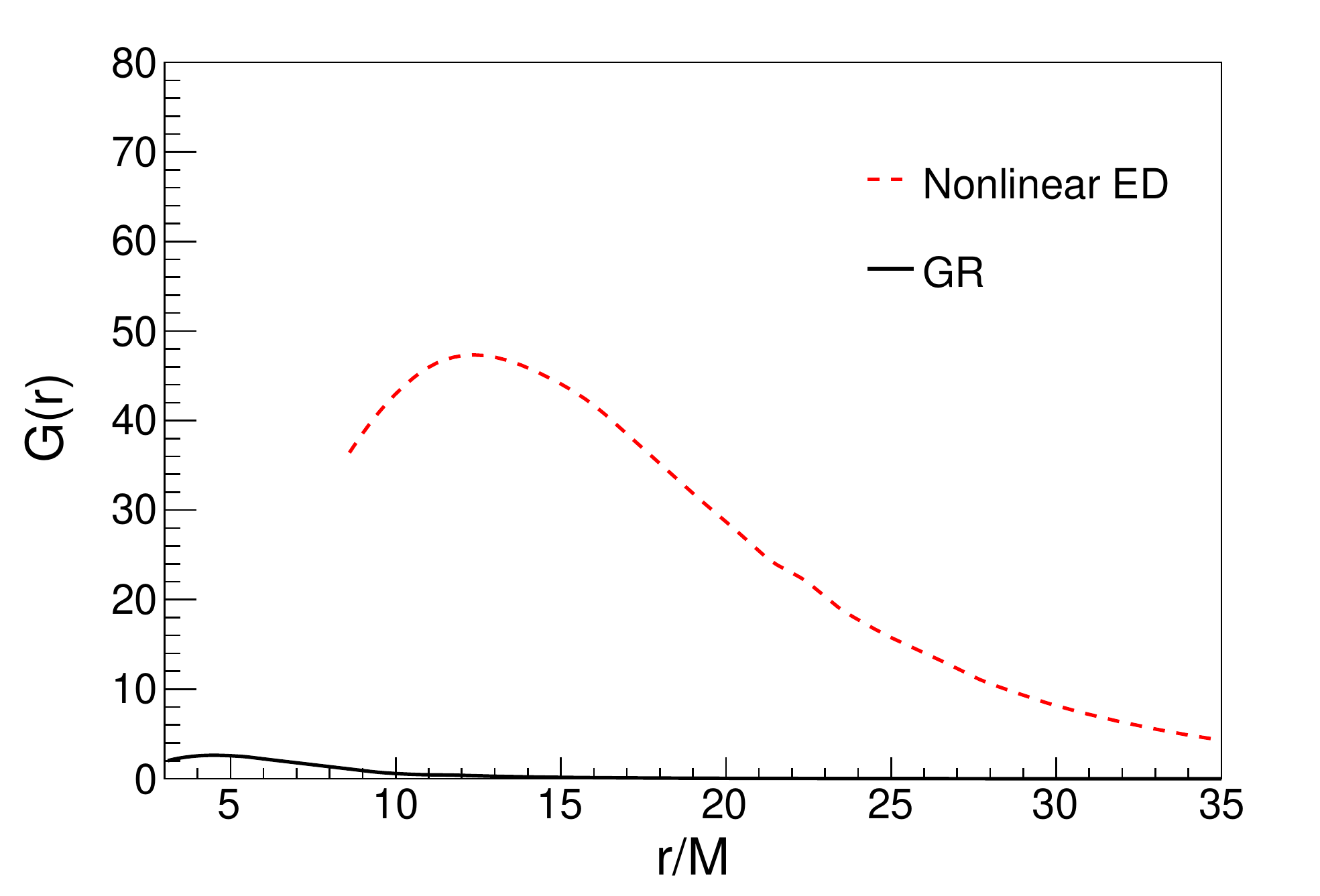}{0.49\textwidth}{(b)}
          }
\caption{Plot of $G(r)$ against $r/M$ for the isothermal disk (a) and temperature gradient model (b). The solid black line is for GR while the dashed red line is for the metric in Eq. (\ref{new_metric}) with $q=1$, $\alpha=1/2$ and $\mu=2$.}
\label{fig:new}
\end{figure*}

Referring to the metric (\ref{new_metric}), the function $G(r)$ entering  into Eq.~(\ref{G(r)Gae}) is plotted in Fig.~\ref{fig:new}~(a) (red dashed line). The EDR, for $q=1$, $\alpha=1/2$ and $\mu=2$, is
\begin{align}
    \frac{dE^{NLE}}{dt}&\simeq 1.6\times 10^{52} \mathrm{erg~ s^{-1}} \,\ , \\
\Pi_{(T=\mathrm{cost})}^{NLE}&\sim 1.06 \,\ . 
\label{piNLE}
\end{align}
which gives a tint deviation from GR. This is due, as in the quintessence model, to the different extension of the disk respect to GR, as arises from Fig.~\ref{fig:new}~(a). The case with $\alpha=1, \mu=2$ is interesting to study because the additional factor in the metric (\ref{new_metric})  is proportional to $r^{-1}$, becoming a redefinition of the mass.

\subsubsection{Temperature gradient model}
As for the previous metrics, using Eq.~(\ref{F(r)}) and considering a temperature varying as $T\propto 2R_{\mathrm{ph}}/r$, it is possible to obtain the function $G(r)$ in Fig.~\ref{fig:new}~(b) for GR (black continuous line) and the charged galileon metric (red dashed line). The enhancement with respect to GR is greater than that in the isothermal model due to the same effect discussed in Sec.~\ref{Quintessence - Temperature gradient}. We obtain:
\begin{align}
    \frac{dE^{NLE}}{dt}&\simeq 1.6\times 10^{51} \mathrm{erg~ s^{-1}} \,\ , \\
    \Pi_{(T\propto 1/r)}^{NLE}&\sim 36 \,\ .
    \label{piNLE1}
\end{align}
As in the case of quintessence, the ED model may provide the energy of the emitted GRB even in the case of a variable temperature. The explanation of the difference between $\Pi_{(T=\mathrm{cost})}^{NLE}$, Eq.~(\ref{piNLE}), and $\Pi_{(T\propto 1/r)}^{NLE}$, Eq.~(\ref{piNLE1}), is identical to the case of quintessence metric (as discussed in Sec.~\ref{Quintessence - Temperature gradient}). In the isothermal model, the $\Pi_{(T=\mathrm{cost})}^{NLE}$ factor, caused by the modification to the neutrino trajectory and space-time geometry, is reduced due to the short disk extension. On the other hand, in the temperature gradient model, we can observe the enhancement without the counter effect associated with the disk extension since $T_0(r)$ quickly drops to lower values with distance.

\subsection{Kerr-Sen black hole}
\begin{figure}
    \centering
    \includegraphics[scale=0.7]{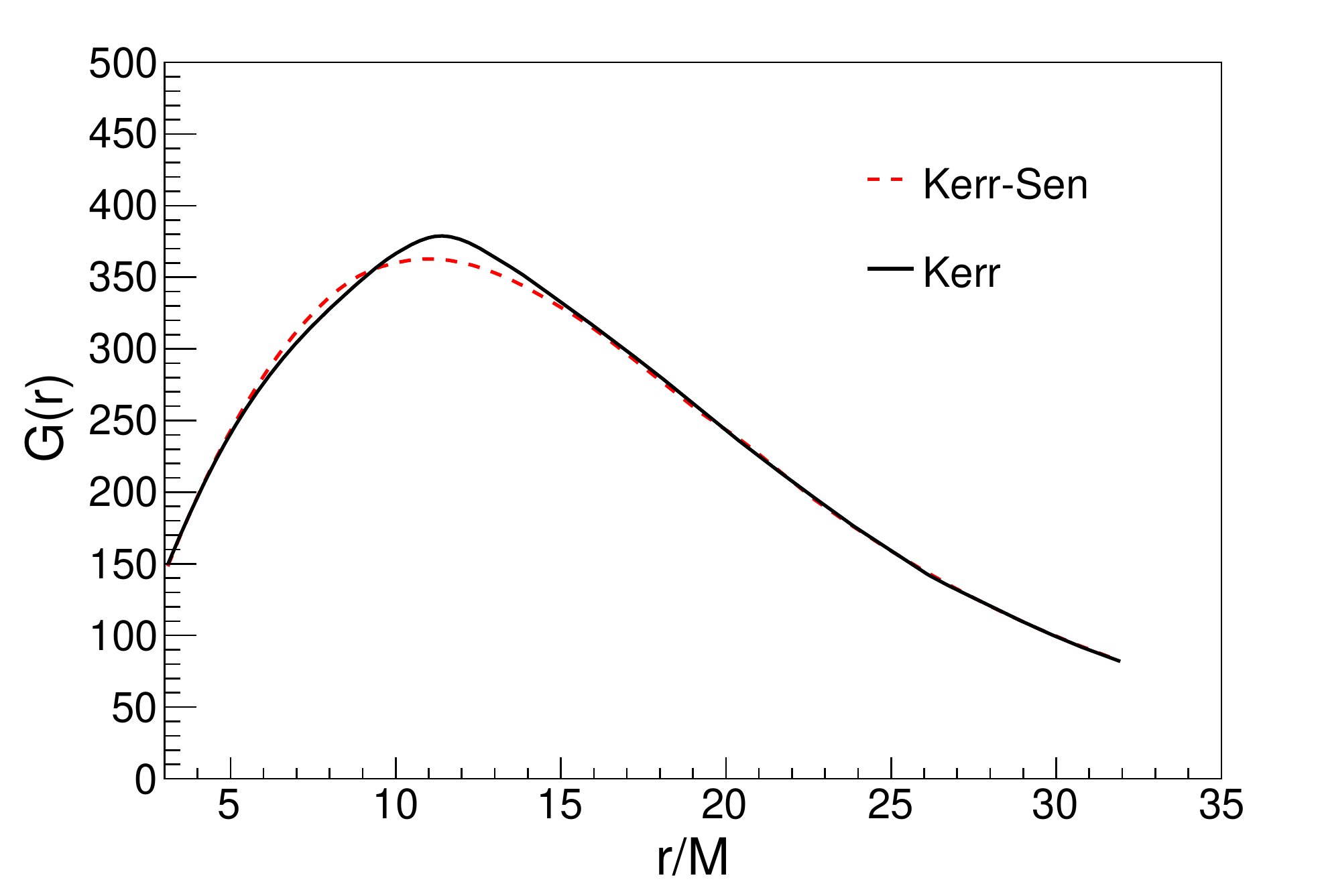}
    \caption{Plot of $G(r)$ against $r/M$ for the isothermal disk. The solid black line is for Kerr metric while the dashed red line is for the Kerr-Sen metric with $a=0.5$ and $Q=0.8$.}
    \label{Kerr-Sen}
\end{figure}

Finally, we consider the Kerr-Sen metric as an example of geometries that, although generalises GR,  do not produce an enhancement of the EDR.
This kind of black hole is a rotating charged black hole solution arising from string theory.
In Ref.~\cite{PhysRevLett.69.1006},  Sen constructed,  in the framework of string theory in 4 dimensions, a rotating charged black hole solution transforming the Kerr black hole solution. The action is given by
\begin{equation}
    S=\int d^4\sqrt{-g}\,\,e^{-\tilde{\phi}}\left(\tilde{R}-\frac{1}{12}\tilde{H}^2-\partial_{\mu}\tilde{\phi}\partial^{\mu}\tilde{\phi}-\frac{\tilde{F}^2}{8}\right) \,\ ,
    \label{Ac:KS}
\end{equation}
where $\tilde{H}$ is the Kalb-Ramond 3-form, $\tilde{F}$ is the gauge filed and ${\tilde \phi}$ is the dilatonic field used for the transformation from the Einstein to string frame, where the quantities are evaluated. From the action  Eq.~(\ref{Ac:KS}) one  obtains a solution of field equations (for an isotropic spherical geometry) given by:
\begin{equation}
\begin{split}
    ds^2&=-\left(1-\frac{2Mr}{\rho^2}\right)dt^2-\frac{4Mra\sin^2\theta}{\rho^2}dtd\phi+\frac{\rho^2}{\Delta}dr^2+\rho^2d\theta^2 +\\
    &+\left(r(r+r_Q)+a^2+\frac{2Mra^2\sin^2\theta}{\rho^2}\right)\sin^2\theta d\phi^2 \,\ ,
    \label{Kerr-sen}
\end{split}
\end{equation}
where $\Delta=r(r+r_Q)-2Mr+a^2$, $r_Q=Q^2/M$, $\rho^2=r(r+r_Q)+a^2\cos^2\theta$ and $a=J^2/M$ with $J$ the angular momentum of BH and $M$ its mass.
For a metric for which $g_{03}\neq 0$, we have to introduce corrections to Eq.~(\ref{rho}), (\ref{Teff}) and (\ref{v^2}). Indeed, it is possible to write that~\cite{Kovacs:2010zp}:
\begin{align}
    T_0(R)&=\frac{T_{\rm{eff}}\left(R,\frac{\pi}{2}\right)}{\gamma}\sqrt{g_{00}\left(R,\frac{\pi}{2}\right)-\frac{g_{03}^2(R,\frac{\pi}{2})}{g_{33}(R,\frac{\pi}{2})}} \,\ , \\
    \frac{v^2}{c^2}&=\frac{g_{33}^2(r,\pi/2)}{g_{03}^2(r,\pi/1)-g_{00}(r,\pi/2)g_{33}(r,\pi/2)}\left(\Omega_K-\omega\right)^2 \,\ , \\
    \Omega_K&=\frac{-g_{03,r}+\sqrt{(g_{03,r})^2-g_{00,r}g_{33,r}}}{g_{33,r}}\Big|_{(r,\pi/2)} \,\ , \\
    \omega&=-\frac{g_{03}(r,\pi/2)}{g_{33}(r,\pi/2)} \,\ , \\
    \rho_\nu&=\sqrt{g_{00}(r,0)g_{22}(r,0)} \,\ ,
\end{align}
where $g_{33}$ and $g_{03}$ are the respective component of the metric defined in Eq.~(\ref{Kerr-sen}).\\
Finally, the equation of the trajectory [Eq.~(\ref{trajectory equation})] becames
\begin{align*}
    &\int\frac{d\theta}{\sqrt{1-(a/\rho_\nu)^2\sin^2\theta}}=\\
    &\int dr'\left[\left(r'(r'+r_Q)\right)\left(\frac{r'(r'+r_Q)}{r_0^2}\left(1+\frac{a^2}{r'(r'+r_Q)}\right)-\left(1-\frac{2M}{r'}+\frac{a^2}{r'(r'+r_Q)}\right)\right)\right]^{-1/2} \,\ .
\end{align*}    
In Fig.~\ref{Kerr-Sen}, we plot the function $G(r)$ (Eq. \ref{G(r)Gae}) for Kerr metric (black continuous line) and Kerr-Sen metric (red dashed line) with $a=0.5$ and $Q=0.8$ for isothermal model. In such a case, the space-time is mostly similar to the Kerr space-time and the enhancement factor is not appreciable, so that the energy emitted is almost equal to that of General Relativity. An identical result has been obtained for the disk temperature gradient model 

\section{Conclusion}
\label{Conclusion}
In this paper, we have investigated the GR effect on the energy deposition rate of $\nu\bar{\nu}\rightarrow e^-e^+$ reaction in
geometries inferred in theories beyond GR. We have used the Eq.~(\ref{trajectory}) to describe the energy deposition rate of neutrinos coming from the equatorial accretion disk and depositing energy along the rotation axis of the star. The bending of the neutrino trajectories and the redshift due to the gravitational field have been taken into consideration, as well. We find that the energy deposition rate function $G(r)$ has a common shape among the various metrics: initially $G(r)$ increases with the distance, reaching
a maximum value, and then, due to the interplay between temperature and redshift effects, decreases with the distance.

We have then calculated the value of the energy deposition rate using Eq.~(\ref{value}) for various BH metrics beyond general relativity, with a characteristic value of the temperature of $10$~MeV and of the angle $\theta=10^\circ$, for which we have found the results in Sec. III. They are close to that of the maximum energy liberated during GRB ($\sim 5\times 10^{52}\mathrm{erg/s}$). Moreover, we find an enhancement of the energy deposition rate with respect to GR for all the considered metrics, up to a factor $\sim 40$ for the Quintessence and Non-linear electrodynamic metrics. The enhancement is caused by the modification, with respect to GR, of the neutrino trajectory and of the $T_{0}(R)$, as described in Sec.~\ref{Formulation}.
This result is very sensitive to the disk temperature. The maximum amount of energy deposited for an isothermal disk with the chosen modified metrics is $\sim 3.0\times 10^{52}\mathrm{erg/s}$ with an enhancement with respect to GR of a factor of $\sim 2$, while for the disk with $T_{\mathrm{eff}}\propto r^{-1}$ is $\sim 1.6\times 10^{51}\mathrm{erg/s}$ with an enhancement with respect to GR of a factor of $\sim 35$.

Finally, we observe that a more accurate estimation
of the total energy deposition requires calculating the energy deposition rate for neutrinos emitted from both the neutron star and the disk, which has to be modelled with temperature profiles coming from neutron star merging simulation. These will be crucial to exactly determine the value of the total energy deposition rate. Such a study is in progress.

 \begin{acknowledgments}
The authors thank O. Just for the discussions and suggestions that improved the paper. The authors thank the referee for the hint in improving the paper.
The work of G.L. and L.M. is supported by the Italian Istituto Nazionale di Fisica Nucleare (INFN) through the ``QGSKY'' project and by Ministero dell'Istruzione, Universit\`a e Ricerca (MIUR).
The computational work has been executed on the IT resources of the ReCaS-Bari data center, which have been made available by two projects financed by the MIUR (Italian Ministry for Education, University and Re-search) in the "PON Ricerca e Competitività 2007-2013" Program: ReCaS (Azione I - Interventi di rafforzamento strutturale, PONa3\_00052, Avviso 254/Ric) and PRISMA (Asse II - Sostegno all'innovazione, PON04a2A)
 \end{acknowledgments}

\newpage
\bibliographystyle{aasjournal.bst}  
\bibliography{sources.bib}

\begin{thebibliography}{}
\expandafter\ifx\csname natexlab\endcsname\relax\def\natexlab#1{#1}\fi
\providecommand{\url}[1]{\href{#1}{#1}}
\providecommand{\dodoi}[1]{doi:~\href{http://doi.org/#1}{\nolinkurl{#1}}}
\providecommand{\doeprint}[1]{\href{http://ascl.net/#1}{\nolinkurl{http://ascl.net/#1}}}
\providecommand{\doarXiv}[1]{\href{https://arxiv.org/abs/#1}{\nolinkurl{https://arxiv.org/abs/#1}}}

\bibitem[{Abbott {et~al.}(2016)Abbott, Abbott, Abbott, Abernathy, Acernese,
  Ackley, Adams, Adams, Addesso, Adhikari, Adya, Affeldt, Agathos, Agatsuma,
  Aggarwal, Aguiar, Aiello, Ain, Ajith, Allen, Allocca, Altin, Anderson,
  Anderson, Arai, Araya, Arceneaux, Areeda, Arnaud, Arun, Ascenzi, Ashton, Ast,
  Aston, Astone, Aufmuth, Aulbert, Babak, Bacon, Bader, Baker, Baldaccini,
  Ballardin, Ballmer, Barayoga, Barclay, Barish, Barker, Barone, Barr,
  Barsotti, Barsuglia, Barta, Bartlett, Bartos, Bassiri, Basti, Batch, Baune,
  Bavigadda, Bazzan, Behnke, Bejger, Bell, Bell, Berger, Bergman, Bergmann,
  Berry, Bersanetti, Bertolini, Betzwieser, Bhagwat, Bhandare, Bilenko,
  Billingsley, Birch, Birney, Birnholtz, Biscans, Bisht, Bitossi, Biwer,
  Bizouard, Blackburn, Blair, Blair, Blair, Bloemen, Bock, Bodiya, Boer,
  Bogaert, Bogan, Bohe, Bojtos, Bond, Bondu, Bonnand, Boom, Bork, Boschi, Bose,
  Bouffanais, Bozzi, Bradaschia, Brady, Braginsky, Branchesi, Brau, Briant,
  Brillet, Brinkmann, Brisson, Brockill, Brooks, Brown, Brown, Brown, Buchanan,
  Buikema, Bulik, Bulten, Buonanno, Buskulic, Buy, Byer, Cadonati, Cagnoli,
  Cahillane, Calder\'on~Bustillo, Callister, Calloni, Camp, Cannon, Cao,
  Capano, Capocasa, Carbognani, Caride, Casanueva~Diaz, Casentini, Caudill,
  Cavagli\`a, Cavalier, Cavalieri, Cella, Cepeda, Cerboni~Baiardi, Cerretani,
  Cesarini, Chakraborty, Chalermsongsak, Chamberlin, Chan, Chao, Charlton,
  Chassande-Mottin, Chen, Chen, Cheng, Chincarini, Chiummo, Cho, Cho, Chow,
  Christensen, Chu, Chua, Chung, Ciani, Clara, Clark, Cleva, Coccia, Cohadon,
  Colla, Collette, Cominsky, Constancio, Conte, Conti, Cook, Corbitt, Cornish,
  Corsi, Cortese, Costa, Coughlin, Coughlin, Coulon, Countryman, Couvares,
  Cowan, Coward, Cowart, Coyne, Coyne, Craig, Creighton, Cripe, Crowder,
  Cumming, Cunningham, Cuoco, Dal~Canton, Danilishin, D'Antonio, Danzmann,
  Darman, Dattilo, Dave, Daveloza, Davier, Davies, Daw, Day, DeBra, Debreczeni,
  Degallaix, De~Laurentis, Del\'eglise, Del~Pozzo, Denker, Dent, Dereli,
  Dergachev, De~Rosa, DeRosa, DeSalvo, Dhurandhar, D\'{\i}az, Di~Fiore,
  Di~Giovanni, Di~Lieto, Di~Pace, Di~Palma, Di~Virgilio, Dojcinoski, Dolique,
  Donovan, Dooley, Doravari, Douglas, Downes, Drago, Drever, Driggers, Du,
  Ducrot, Dwyer, Edo, Edwards, Effler, Eggenstein, Ehrens, Eichholz,
  Eikenberry, Engels, Essick, Etzel, Evans, Evans, Everett, Factourovich,
  Fafone, Fair, Fairhurst, Fan, Fang, Farinon, Farr, Farr, Favata, Fays,
  Fehrmann, Fejer, Ferrante, Ferreira, Ferrini, Fidecaro, Fiori, Fiorucci,
  Fisher, Flaminio, Fletcher, Fournier, Franco, Frasca, Frasconi, Frei, Freise,
  Frey, Frey, Fricke, Fritschel, Frolov, Fulda, Fyffe, Gabbard, Gair,
  Gammaitoni, Gaonkar, Garufi, Gatto, Gaur, Gehrels, Gemme, Gendre, Genin,
  Gennai, George, Gergely, Germain, Ghosh, Ghosh, Ghosh, Giaime, Giardina,
  Giazotto, Gill, Glaefke, Goetz, Goetz, Gondan, Gonz\'alez, Gonzalez~Castro,
  Gopakumar, Gordon, Gorodetsky, Gossan, Gosselin, Gouaty, Graef, Graff,
  Granata, Grant, Gras, Gray, Greco, Green, Groot, Grote, Grunewald, Guidi,
  Guo, Gupta, Gupta, Gushwa, Gustafson, Gustafson, Hacker, Hall, Hall, Hammond,
  Haney, Hanke, Hanks, Hanna, Hannam, Hanson, Hardwick, Harms, Harry, Harry,
  Hart, Hartman, Haster, Haughian, Healy, Heidmann, Heintze, Heitmann, Hello,
  Hemming, Hendry, Heng, Hennig, Heptonstall, Heurs, Hild, Hoak, Hodge, Hofman,
  Hollitt, Holt, Holz, Hopkins, Hosken, Hough, Houston, Howell, Hu, Huang,
  Huerta, Huet, Hughey, Husa, Huttner, Huynh-Dinh, Idrisy, Indik, Ingram, Inta,
  Isa, Isac, Isi, Islas, Isogai, Iyer, Izumi, Jacqmin, Jang, Jani, Jaranowski,
  Jawahar, Jim\'enez-Forteza, Johnson, Johnson-McDaniel, Jones, Jones, Jonker,
  Ju, Haris, Kalaghatgi, Kalogera, Kandhasamy, Kang, Kanner, Karki, Kasprzack,
  Katsavounidis, Katzman, Kaufer, Kaur, Kawabe, Kawazoe, K\'ef\'elian, Kehl,
  Keitel, Kelley, Kells, Kennedy, Key, Khalaidovski, Khalili, Khan, Khan, Khan,
  Khazanov, Kijbunchoo, Kim, Kim, Kim, Kim, Kim, Kim, King, King, Kinzel,
  Kissel, Kleybolte, Klimenko, Koehlenbeck, Kokeyama, Koley, Kondrashov,
  Kontos, Korobko, Korth, Kowalska, Kozak, Kringel, Krishnan, Kr\'olak,
  Krueger, Kuehn, Kumar, Kuo, Kutynia, Lackey, Landry, Lange, Lantz, Lasky,
  Lazzarini, Lazzaro, Leaci, Leavey, Lebigot, Lee, Lee, Lee, Lee, Lenon,
  Leonardi, Leong, Leroy, Letendre, Levin, Levine, Li, Libson, Littenberg,
  Lockerbie, Logue, Lombardi, London, Lord, Lorenzini, Loriette, Lormand,
  Losurdo, Lough, Lousto, Lovelace, L\"uck, Lundgren, Luo, Lynch, Ma,
  MacDonald, Machenschalk, MacInnis, Macleod, Maga\~na Sandoval, Magee,
  Mageswaran, Majorana, Maksimovic, Malvezzi, Man, Mandel, Mandic, Mangano,
  Mansell, Manske, Mantovani, Marchesoni, Marion, M\'arka, M\'arka, Markosyan,
  Maros, Martelli, Martellini, Martin, Martin, Martynov, Marx, Mason, Masserot,
  Massinger, Masso-Reid, Matichard, Matone, Mavalvala, Mazumder, Mazzolo,
  McCarthy, McClelland, McCormick, McGuire, McIntyre, McIver, McManus,
  McWilliams, Meacher, Meadors, Meidam, Melatos, Mendell, Mendoza-Gandara,
  Mercer, Merilh, Merzougui, Meshkov, Messenger, Messick, Meyers, Mezzani,
  Miao, Michel, Middleton, Mikhailov, Milano, Miller, Millhouse, Minenkov,
  Ming, Mirshekari, Mishra, Mitra, Mitrofanov, Mitselmakher, Mittleman, Moggi,
  Mohan, Mohapatra, Montani, Moore, Moore, Moraru, Moreno, Morriss, Mossavi,
  Mours, Mow-Lowry, Mueller, Mueller, Muir, Mukherjee, Mukherjee, Mukherjee,
  Mukund, Mullavey, Munch, Murphy, Murray, Mytidis, Nardecchia, Naticchioni,
  Nayak, Necula, Nedkova, Nelemans, Neri, Neunzert, Newton, Nguyen, Nielsen,
  Nissanke, Nitz, Nocera, Nolting, Normandin, Nuttall, Oberling, Ochsner,
  O'Dell, Oelker, Ogin, Oh, Oh, Ohme, Oliver, Oppermann, Oram, O'Reilly,
  O'Shaughnessy, Ottaway, Ottens, Overmier, Owen, Pai, Pai, Palamos, Palashov,
  Palomba, Pal-Singh, Pan, Pan, Pankow, Pannarale, Pant, Paoletti, Paoli, Papa,
  Paris, Parker, Pascucci, Pasqualetti, Passaquieti, Passuello, Patricelli,
  Patrick, Pearlstone, Pedraza, Pedurand, Pekowsky, Pele, Penn, Perreca,
  Pfeiffer, Phelps, Piccinni, Pichot, Piergiovanni, Pierro, Pillant, Pinard,
  Pinto, Pitkin, Poggiani, Popolizio, Post, Powell, Prasad, Predoi,
  Premachandra, Prestegard, Price, Prijatelj, Principe, Privitera, Prix, Prodi,
  Prokhorov, Puncken, Punturo, Puppo, P\"urrer, Qi, Qin, Quetschke, Quintero,
  Quitzow-James, Raab, Rabeling, Radkins, Raffai, Raja, Rakhmanov, Rapagnani,
  Raymond, Razzano, Re, Read, Reed, Regimbau, Rei, Reid, Reitze, Rew, Reyes,
  Ricci, Riles, Robertson, Robie, Robinet, Rocchi, Rolland, Rollins, Roma,
  Romano, Romanov, Romie, Rosi\ifmmode~\acute{n}\else \'{n}\fi{}ska, Rowan,
  R\"udiger, Ruggi, Ryan, Sachdev, Sadecki, Sadeghian, Salconi, Saleem, Salemi,
  Samajdar, Sammut, Sanchez, Sandberg, Sandeen, Sanders, Sassolas,
  Sathyaprakash, Saulson, Sauter, Savage, Sawadsky, Schale, Schilling, Schmidt,
  Schmidt, Schnabel, Schofield, Sch\"onbeck, Schreiber, Schuette, Schutz,
  Scott, Scott, Sellers, Sengupta, Sentenac, Sequino, Sergeev, Serna,
  Setyawati, Sevigny, Shaddock, Shah, Shahriar, Shaltev, Shao, Shapiro,
  Shawhan, Sheperd, Shoemaker, Shoemaker, Siellez, Siemens, Sigg, Silva,
  Simakov, Singer, Singer, Singh, Singh, Singhal, Sintes, Slagmolen, Smith,
  Smith, Smith, Son, Sorazu, Sorrentino, Souradeep, Srivastava, Staley,
  Steinke, Steinlechner, Steinlechner, Steinmeyer, Stephens, Stone, Strain,
  Straniero, Stratta, Strauss, Strigin, Sturani, Stuver, Summerscales, Sun,
  Sutton, Swinkels, Szczepa\ifmmode~\acute{n}\else \'{n}\fi{}czyk, Tacca,
  Talukder, Tanner, T\'apai, Tarabrin, Taracchini, Taylor, Theeg,
  Thirugnanasambandam, Thomas, Thomas, Thomas, Thorne, Thorne, Thrane, Tiwari,
  Tiwari, Tokmakov, Tomlinson, Tonelli, Torres, Torrie, T\"oyr\"a, Travasso,
  Traylor, Trifir\`o, Tringali, Trozzo, Tse, Turconi, Tuyenbayev, Ugolini,
  Unnikrishnan, Urban, Usman, Vahlbruch, Vajente, Valdes, Vallisneri, van
  Bakel, van Beuzekom, van~den Brand, Van Den~Broeck, Vander-Hyde, van~der
  Schaaf, van Heijningen, van Veggel, Vardaro, Vass, Vas\'uth, Vaulin, Vecchio,
  Vedovato, Veitch, Veitch, Venkateswara, Verkindt, Vetrano, Vicer\'e,
  Vinciguerra, Vine, Vinet, Vitale, Vo, Vocca, Vorvick, Voss, Vousden,
  Vyatchanin, Wade, Wade, Wade, Walker, Wallace, Walsh, Wang, Wang, Wang, Wang,
  Wang, Ward, Warner, Was, Weaver, Wei, Weinert, Weinstein, Weiss, Welborn,
  Wen, We\ss{}els, Westphal, Wette, Whelan, White, Whiting, Williams, Williams,
  Williamson, Willis, Willke, Wimmer, Winkler, Wipf, Wittel, Woan, Worden,
  Wright, Wu, Yablon, Yam, Yamamoto, Yancey, Yap, Yu, Yvert,
  Zadro\ifmmode~\dot{z}\else \.{z}\fi{}ny, Zangrando, Zanolin, Zendri, Zevin,
  Zhang, Zhang, Zhang, Zhang, Zhao, Zhou, Zhou, Zhu, Zucker, Zuraw, Zweizig,
  Boyle, Campanelli, Hemberger, Kidder, Ossokine, Scheel, Szilagyi, Teukolsky,
  \& Zlochower}]{PhysRevLett.116.221101}
Abbott, B.~P., Abbott, R., Abbott, T.~D., {et~al.} 2016, Phys. Rev. Lett., 116,
  221101, \dodoi{10.1103/PhysRevLett.116.221101}

\bibitem[{Abbott {et~al.}(2019)}]{LIGOScientific:2019fpa}
Abbott, B.~P., {et~al.} 2019, Phys. Rev. D, 100, 104036,
  \dodoi{10.1103/PhysRevD.100.104036}

\bibitem[{Akiyama {et~al.}(2019{\natexlab{a}})}]{EventHorizonTelescope:2019dse}
Akiyama, K., {et~al.} 2019{\natexlab{a}}, Astrophys. J. Lett., 875, L1,
  \dodoi{10.3847/2041-8213/ab0ec7}

\bibitem[{Akiyama {et~al.}(2019{\natexlab{b}})}]{EventHorizonTelescope:2019ths}
---. 2019{\natexlab{b}}, Astrophys. J. Lett., 875, L4,
  \dodoi{10.3847/2041-8213/ab0e85}

\bibitem[{Amendola \& Tsujikawa(2015)}]{Amendola:2015ksp}
Amendola, L., \& Tsujikawa, S. 2015, {Dark Energy}: {Theory and Observations}
  (Cambridge University Press)

\bibitem[{Asano \& Fukuyama(2000)}]{Asano:2000ib}
Asano, K., \& Fukuyama, T. 2000, Astrophys. J., 531, 949,
  \dodoi{10.1086/308513}

\bibitem[{Asano \& Fukuyama(2001)}]{Asano:2000dq}
---. 2001, Astrophys. J., 546, 1019, \dodoi{10.1086/318312}

\bibitem[{Babichev {et~al.}(2015)Babichev, Charmousis, \&
  Hassaine}]{Babichev:2015rva}
Babichev, E., Charmousis, C., \& Hassaine, M. 2015, JCAP, 05, 031,
  \dodoi{10.1088/1475-7516/2015/05/031}

\bibitem[{Baker {et~al.}(2017)Baker, Bellini, Ferreira, Lagos, Noller, \&
  Sawicki}]{Baker:2017hug}
Baker, T., Bellini, E., Ferreira, P.~G., {et~al.} 2017, Phys. Rev. Lett., 119,
  251301, \dodoi{10.1103/PhysRevLett.119.251301}

\bibitem[{Birkl {et~al.}(2007)Birkl, Aloy, Janka, \& Mueller}]{Birkl:2006mu}
Birkl, R., Aloy, M.-A., Janka, H.-T., \& Mueller, E. 2007, Astron. Astrophys.,
  463, 51, \dodoi{10.1051/0004-6361:20066293}

\bibitem[{Brans \& Dicke(1961)}]{Brans:1961sx}
Brans, C., \& Dicke, R. 1961, Phys. Rev., 124, 925,
  \dodoi{10.1103/PhysRev.124.925}

\bibitem[{Bronnikov(2017)}]{Bronnikov:2017tnz}
Bronnikov, K.~A. 2017, Phys. Rev. D, 96, 128501,
  \dodoi{10.1103/PhysRevD.96.128501}

\bibitem[{Capozziello \& De~Laurentis(2011)}]{Capozziello:2011et}
Capozziello, S., \& De~Laurentis, M. 2011, Phys. Rept., 509, 167,
  \dodoi{10.1016/j.physrep.2011.09.003}

\bibitem[{Capozziello \& Lambiase(1999)}]{cosmo9}
Capozziello, S., \& Lambiase, G. 1999, Gen. Rel. Grav., 31, 1005,
  \dodoi{10.1023/A:1026631531309}

\bibitem[{Capozziello \& Lambiase(2000)}]{cosmo8}
---. 2000, Gen. Rel. Grav., 32, 295, \dodoi{10.1023/A:1001935510837}

\bibitem[{Carson \& Yagi(2020)}]{Carson:2019rda}
Carson, Z., \& Yagi, K. 2020, Class. Quant. Grav., 37, 02LT01,
  \dodoi{10.1088/1361-6382/ab5c9a}

\bibitem[{Chan {et~al.}(2009)Chan, Cheng, Harko, Lau, Lin, Suen, \&
  Tian}]{Chan:2009mw}
Chan, T., Cheng, K., Harko, T., {et~al.} 2009, Astrophys. J., 695, 732,
  \dodoi{10.1088/0004-637X/695/1/732}

\bibitem[{Chen {et~al.}(2008)Chen, Wang, \& Su}]{Chen:2008ra}
Chen, S., Wang, B., \& Su, R. 2008, Phys. Rev. D, 77, 124011,
  \dodoi{10.1103/PhysRevD.77.124011}

\bibitem[{Clifton {et~al.}(2012)Clifton, Ferreira, Padilla, \&
  Skordis}]{Clifton:2011jh}
Clifton, T., Ferreira, P.~G., Padilla, A., \& Skordis, C. 2012, Phys. Rept.,
  513, 1, \dodoi{10.1016/j.physrep.2012.01.001}

\bibitem[{Cooperstein {et~al.}(1986)Cooperstein, van~den Horn, \& Baron}]{Co86}
Cooperstein, J., van~den Horn, L., \& Baron, E.~A. 1986, ApJ, 309, 653,
  \dodoi{10.1086/164633}

\bibitem[{Cooperstein {et~al.}(1987)Cooperstein, van~den Horn, \& Baron}]{Co87}
Cooperstein, J., van~den Horn, L.~J., \& Baron, E.~A. 1987, ApJ, 321, L129,
  \dodoi{10.1086/185019}

\bibitem[{Creminelli \& Vernizzi(2017)}]{Creminelli:2017sry}
Creminelli, P., \& Vernizzi, F. 2017, Phys. Rev. Lett., 119, 251302,
  \dodoi{10.1103/PhysRevLett.119.251302}

\bibitem[{Dainotti {et~al.}(2022)Dainotti, De~Simone, Schiavone, Montani,
  Rinaldi, Lambiase, Bogdan, \& Ugale}]{Dainotti:2022bzg}
Dainotti, M.~G., De~Simone, B., Schiavone, T., {et~al.} 2022, Galaxies, 10, 24,
  \dodoi{10.3390/galaxies10010024}

\bibitem[{De~Felice \& Tsujikawa(2010)}]{DeFelice:2010aj}
De~Felice, A., \& Tsujikawa, S. 2010, Living Rev. Rel., 13, 3,
  \dodoi{10.12942/lrr-2010-3}

\bibitem[{{Eichler} {et~al.}(1989){Eichler}, {Livio}, {Piran}, \&
  {Schramm}}]{1989Natur.340..126E}
{Eichler}, D., {Livio}, M., {Piran}, T., \& {Schramm}, D.~N. 1989, \nat, 340,
  126, \dodoi{10.1038/340126a0}

\bibitem[{Ezquiaga \& Zumalac\'arregui(2017)}]{Ezquiaga:2017ekz}
Ezquiaga, J.~M., \& Zumalac\'arregui, M. 2017, Phys. Rev. Lett., 119, 251304,
  \dodoi{10.1103/PhysRevLett.119.251304}

\bibitem[{Ezquiaga \& Zumalac\'arregui(2018)}]{Ezquiaga:2018btd}
---. 2018, Front. Astron. Space Sci., 5, 44, \dodoi{10.3389/fspas.2018.00044}

\bibitem[{Fan \& Wang(2016)}]{Fan:2016hvf}
Fan, Z.-Y., \& Wang, X. 2016, Phys. Rev. D, 94, 124027,
  \dodoi{10.1103/PhysRevD.94.124027}

\bibitem[{{Foucart} {et~al.}(2020){Foucart}, {Duez}, {Hebert}, {Kidder},
  {Pfeiffer}, \& {Scheel}}]{2020ApJ...902L..27F}
{Foucart}, F., {Duez}, M.~D., {Hebert}, F., {et~al.} 2020, ApJ, 902, L27,
  \dodoi{10.3847/2041-8213/abbb87}

\bibitem[{Foucart {et~al.}(2018)Foucart, Duez, Kidder, Nguyen, Pfeiffer, \&
  Scheel}]{PhysRevD.98.063007}
Foucart, F., Duez, M.~D., Kidder, L.~E., {et~al.} 2018, Phys. Rev. D, 98,
  063007, \dodoi{10.1103/PhysRevD.98.063007}

\bibitem[{Fujibayashi {et~al.}(2017)Fujibayashi, Sekiguchi, Kiuchi, \&
  Shibata}]{Fujibayashi_2017}
Fujibayashi, S., Sekiguchi, Y., Kiuchi, K., \& Shibata, M. 2017, The
  Astrophysical Journal, 846, 114, \dodoi{10.3847/1538-4357/aa8039}

\bibitem[{Gnocchi {et~al.}(2019)Gnocchi, Maselli, Abdelsalhin, Giacobbo, \&
  Mapelli}]{Gnocchi:2019jzp}
Gnocchi, G., Maselli, A., Abdelsalhin, T., Giacobbo, N., \& Mapelli, M. 2019,
  Phys. Rev. D, 100, 064024, \dodoi{10.1103/PhysRevD.100.064024}

\bibitem[{Goodman {et~al.}(1987)Goodman, Dar, \& Nussinov}]{Goodman:1986we}
Goodman, J., Dar, A., \& Nussinov, S. 1987, Astrophys. J. Lett., 314, L7,
  \dodoi{10.1086/184840}

\bibitem[{Harikae {et~al.}(2010)Harikae, Kotake, Takiwaki, \& ichiro
  Sekiguchi}]{Harikae_2010}
Harikae, S., Kotake, K., Takiwaki, T., \& ichiro Sekiguchi, Y. 2010, The
  Astrophysical Journal, 720, 614, \dodoi{10.1088/0004-637x/720/1/614}

\bibitem[{{Jaroszynski}(1993)}]{1993AcA....43..183J}
{Jaroszynski}, M. 1993, Acta Astron., 43, 183

\bibitem[{Just {et~al.}(2016)Just, Obergaulinger, Janka, Bauswein, \&
  Schwarz}]{Just:2015dba}
Just, O., Obergaulinger, M., Janka, H.~T., Bauswein, A., \& Schwarz, N. 2016,
  Astrophys. J. Lett., 816, L30, \dodoi{10.3847/2041-8205/816/2/L30}

\bibitem[{{Kawanaka} \& {Mineshige}(2007)}]{2007ApJ...662.1156K}
{Kawanaka}, N., \& {Mineshige}, S. 2007, \apj, 662, 1156,
  \dodoi{10.1086/517985}

\bibitem[{Kiselev(2003)}]{Kiselev:2002dx}
Kiselev, V. 2003, Class. Quant. Grav., 20, 1187,
  \dodoi{10.1088/0264-9381/20/6/310}

\bibitem[{Kokkotas {et~al.}(2017)Kokkotas, Konoplya, \&
  Zhidenko}]{Kokkotas:2017zwt}
Kokkotas, K., Konoplya, R., \& Zhidenko, A. 2017, Phys. Rev. D, 96, 064007,
  \dodoi{10.1103/PhysRevD.96.064007}

\bibitem[{Kovacs {et~al.}(2010)Kovacs, Cheng, \& Harko}]{Kovacs:2009dv}
Kovacs, Z., Cheng, K., \& Harko, T. 2010, Mon. Not. Roy. Astron. Soc., 402,
  1714, \dodoi{10.1111/j.1365-2966.2009.15986.x}

\bibitem[{Kovacs {et~al.}(2011)Kovacs, Cheng, \& Harko}]{Kovacs:2010zp}
Kovacs, Z., Cheng, K.~S., \& Harko, T. 2011, Mon. Not. Roy. Astron. Soc., 411,
  1503, \dodoi{10.1111/j.1365-2966.2010.17784.x}

\bibitem[{Lambiase \& Mastrototaro(2020)}]{Lambiase:2020iul}
Lambiase, G., \& Mastrototaro, L. 2020, ApJ., 904, 1,
  \dodoi{10.3847/1538-4357/abba2c}

\bibitem[{Lambiase \& Mastrototaro(2021)}]{Lambiase:2020pkc}
---. 2021, Eur. Phys. J. C, 81, 932, \dodoi{10.1140/epjc/s10052-021-09732-2}

\bibitem[{Lima {et~al.}(2021)Lima, Crispino, Cunha, \& Herdeiro}]{Lima:2021las}
Lima, Junior., H. C.~D., Crispino, L. C.~B., Cunha, P. V.~P., \& Herdeiro, C.
  A.~R. 2021, Phys. Rev. D, 103, 084040, \dodoi{10.1103/PhysRevD.103.084040}

\bibitem[{{Liu} {et~al.}(2015){Liu}, {Gu}, {Kawanaka}, \&
  {Li}}]{2015ApJ...805...37L}
{Liu}, T., {Gu}, W.-M., {Kawanaka}, N., \& {Li}, A. 2015, \apj, 805, 37,
  \dodoi{10.1088/0004-637X/805/1/37}

\bibitem[{Liu {et~al.}(2007)Liu, Gu, Xue, \& Lu}]{Liu:2007bca}
Liu, T., Gu, W.-M., Xue, L., \& Lu, J.-F. 2007, Astrophys. J., 661, 1025,
  \dodoi{10.1086/513689}

\bibitem[{Mallick {et~al.}(2013)Mallick, Bhattacharyya, Ghosh, \&
  Raha}]{Bhattacharyya:2009nm}
Mallick, R., Bhattacharyya, A., Ghosh, S.~K., \& Raha, S. 2013, Int. J. Mod.
  Phys. E, 22, 1350008, \dodoi{10.1142/S0218301313500080}

\bibitem[{Mallick \& Majumder(2009)}]{Mallick:2008iv}
Mallick, R., \& Majumder, S. 2009, Phys. Rev. D, 79, 023001,
  \dodoi{10.1103/PhysRevD.79.023001}

\bibitem[{Matteo {et~al.}(2002)Matteo, Perna, \& Narayan}]{Di_Matteo_2002}
Matteo, T.~D., Perna, R., \& Narayan, R. 2002, The Astrophysical Journal, 579,
  706, \dodoi{10.1086/342832}

\bibitem[{Mineshige {et~al.}(1999)Mineshige, Yonehara, \&
  Kawaguchi}]{10.1143/PTPS.136.235}
Mineshige, S., Yonehara, A., \& Kawaguchi, T. 1999, Progress of Theoretical
  Physics Supplement, 136, 235, \dodoi{10.1143/PTPS.136.235}

\bibitem[{Mukherjee \& Chakraborty(2018)}]{Mukherjee:2017fqz}
Mukherjee, S., \& Chakraborty, S. 2018, Phys. Rev. D, 97, 124007,
  \dodoi{10.1103/PhysRevD.97.124007}

\bibitem[{Narayan {et~al.}(1998)Narayan, Mahadevan, \&
  Quataert}]{Narayan:1998ft}
Narayan, R., Mahadevan, R., \& Quataert, E. 1998.
\newblock \doarXiv{astro-ph/9803141}

\bibitem[{Nojiri \& Odintsov(2017)}]{Nojiri:2017kex}
Nojiri, S., \& Odintsov, S.~D. 2017, Phys. Rev. D, 96, 104008,
  \dodoi{10.1103/PhysRevD.96.104008}

\bibitem[{Okyay \& \"Ovg\"un(2022)}]{Okyay:2021nnh}
Okyay, M., \& \"Ovg\"un, A. 2022, JCAP, 01, 009,
  \dodoi{10.1088/1475-7516/2022/01/009}

\bibitem[{Popham {et~al.}(1999)Popham, Woosley, \& Fryer}]{Popham_1999}
Popham, R., Woosley, S.~E., \& Fryer, C. 1999, The Astrophysical Journal, 518,
  356, \dodoi{10.1086/307259}

\bibitem[{Prasanna \& Goswami(2002)}]{Prasanna:2001ie}
Prasanna, A., \& Goswami, S. 2002, Phys. Lett. B, 526, 27,
  \dodoi{10.1016/S0370-2693(01)01470-8}

\bibitem[{Riess {et~al.}(1998)}]{riess}
Riess, A.~G., {et~al.} 1998, Astron. J., 116, 1009, \dodoi{10.1086/300499}

\bibitem[{Rodrigues \& Junior(2017)}]{PhysRevD.96.128502}
Rodrigues, M.~E., \& Junior, E. L.~B. 2017, Phys. Rev. D, 96, 128502,
  \dodoi{10.1103/PhysRevD.96.128502}

\bibitem[{{Ruffert} \& {Janka}(1999)}]{1999A&A...344..573R}
{Ruffert}, M., \& {Janka}, H.~T. 1999, Astron. Astrophys., 344, 573.
\newblock \doarXiv{astro-ph/9809280}

\bibitem[{Sakstein \& Jain(2017)}]{Sakstein:2017xjx}
Sakstein, J., \& Jain, B. 2017, Phys. Rev. Lett., 119, 251303,
  \dodoi{10.1103/PhysRevLett.119.251303}

\bibitem[{Salmonson \& Wilson(1999)}]{Salmonson:1999es}
Salmonson, J.~D., \& Wilson, J.~R. 1999, Astrophys. J., 517, 859,
  \dodoi{10.1086/307232}

\bibitem[{Salmonson \& Wilson(2001)}]{Salmonson:2001tz}
---. 2001, Astrophys. J., 561, 950, \dodoi{10.1086/323319}

\bibitem[{Sen(1992)}]{PhysRevLett.69.1006}
Sen, A. 1992, Phys. Rev. Lett., 69, 1006, \dodoi{10.1103/PhysRevLett.69.1006}

\bibitem[{Sotiriou \& Faraoni(2010)}]{odi}
Sotiriou, T.~P., \& Faraoni, V. 2010, Rev. Mod. Phys., 82, 451,
  \dodoi{10.1103/RevModPhys.82.451}

\bibitem[{Stuchl\'\i{}k \& Schee(2019)}]{Stuchlik:2019uvf}
Stuchl\'\i{}k, Z., \& Schee, J. 2019, Eur. Phys. J. C, 79, 44,
  \dodoi{10.1140/epjc/s10052-019-6543-8}

\bibitem[{Stuchl{\'{\i}}k {et~al.}(2019)Stuchl{\'{\i}}k, Schee, \&
  Ovchinnikov}]{Stuchlik2019}
Stuchl{\'{\i}}k, Z., Schee, J., \& Ovchinnikov, D. 2019, The Astrophysical
  Journal, 887, 145, \dodoi{10.3847/1538-4357/ab55d5}

\bibitem[{Toshmatov {et~al.}(2018{\natexlab{a}})Toshmatov, Stuchl\'\i{}k, \&
  Ahmedov}]{Toshmatov:2018ell}
Toshmatov, B., Stuchl\'\i{}k, Z., \& Ahmedov, B. 2018{\natexlab{a}}, Phys. Rev.
  D, 98, 085021, \dodoi{10.1103/PhysRevD.98.085021}

\bibitem[{Toshmatov {et~al.}(2018{\natexlab{b}})Toshmatov, Stuchl\'{\i}k, \&
  Ahmedov}]{PhysRevD.98.028501}
Toshmatov, B., Stuchl\'{\i}k, Z. c.~v., \& Ahmedov, B. 2018{\natexlab{b}},
  Phys. Rev. D, 98, 028501, \dodoi{10.1103/PhysRevD.98.028501}

\bibitem[{Tretyakova \& Adyev(2016)}]{Tretyakova:2016ale}
Tretyakova, D.~A., \& Adyev, T.~M. 2016.
\newblock \doarXiv{1610.07300}

\bibitem[{Wu \& Zhang(2021)}]{Wu:2021pgf}
Wu, X., \& Zhang, X. 2021.
\newblock \doarXiv{2112.11066}

\bibitem[{Xavier {et~al.}(2020)Xavier, Cunha, Crispino, \&
  Herdeiro}]{Xavier:2020egv}
Xavier, S. V. M. C.~B., Cunha, P. V.~P., Crispino, L. C.~B., \& Herdeiro, C.
  A.~R. 2020, Int. J. Mod. Phys. D, 29, 2041005,
  \dodoi{10.1142/S0218271820410059}

\bibitem[{Zalamea \& Beloborodov(2009)}]{doi:10.1063/1.3155863}
Zalamea, I., \& Beloborodov, A.~M. 2009, AIP Conference Proceedings, 1133, 121,
  \dodoi{10.1063/1.3155863}

\bibitem[{Zeng \& Zhang(2020)}]{Zeng:2020vsj}
Zeng, X.-X., \& Zhang, H.-Q. 2020, Eur. Phys. J. C, 80, 1058,
  \dodoi{10.1140/epjc/s10052-020-08656-7}

\end{thebibliography}

\end{document}